\documentclass[a4paper]{article}
\usepackage{geometry}
\usepackage{amsmath}
\usepackage{amssymb}
\usepackage{indentfirst}
\usepackage{mathdots}
\usepackage{cite}
\usepackage[colorlinks, linkcolor=blue, citecolor=blue]{hyperref}
\usepackage{mathrsfs}
\usepackage{booktabs}
\usepackage{enumitem}
\usepackage{authblk}
\usepackage{amsfonts,ntheorem}
\usepackage{makecell}
\numberwithin{equation}{section}

\makeatletter

\newcommand{\Rmnum}[1]{\expandafter\@slowromancap\romannumeral #1@}
\makeatother

\newtheorem{theorem}{Theorem}[section]
\newtheorem{lemma}[theorem]{Lemma}
\newtheorem{corollary}[theorem]{Corollary}
\newtheorem{proposition}[theorem]{Proposition}

{ 
\theoremstyle{plain}
\theorembodyfont{\normalfont} 

\newtheorem{definition}[theorem]{Definition}

\newtheorem{remark}[theorem]{Remark}

}

\newenvironment{proof}{\noindent{\textbf{\emph{Proof.}}}}

\allowdisplaybreaks[4]

\begin {document}
\title{\Large Intersections of linear codes and related MDS codes with new Galois hulls}
\author[1]{{\small{Meng Cao}} \thanks{E-mail address: mengcaomath@126.com}}
\author[2]{{\small{Jing Yang}} \thanks{E-mail address: y-j@tsinghua.edu.cn}}
\affil[1]{\footnotesize{Yanqi Lake Beijing Institute of Mathematical Sciences and Applications, Beijing, 101408, China} }
\affil[2]{\footnotesize{Department of Mathematical Sciences, Tsinghua University, Beijing 100084, China} }

\date{}
\maketitle
\vspace{-15pt}
{\linespread{1.4}{

\begin{abstract}
Let $\mathrm{SLAut}(\mathbb{F}_{q}^{n})$ denote the group of all semilinear isometries on $\mathbb{F}_{q}^{n}$, where $q=p^{e}$ is a prime power.
In this paper, we investigate general properties of linear codes associated with $\sigma$ duals
for $\sigma\in\mathrm{SLAut}(\mathbb{F}_{q}^{n})$.
We show that the dimension of the intersection of two linear codes can be determined by generator matrices of such codes and their $\sigma$ duals.
We also show that the dimension of $\sigma$ hull of a linear code can be determined by a generator matrix of it or its $\sigma$ dual.
We give a characterization on $\sigma$ dual and $\sigma$ hull of a matrix-product code.
We also investigate the intersection of a pair of matrix-product codes.
We provide a necessary and sufficient condition under which any codeword of a generalized Reed-Solomon (GRS) code or an extended GRS code is contained in
its $\sigma$ dual.
As an application, we construct eleven families of $q$-ary MDS codes with new $\ell$-Galois hulls satisfying $2(e-\ell)\mid e$,
which are not covered by the latest papers by Cao (IEEE Trans. Inf. Theory 67(12), 7964-7984, 2021)
and by Fang \emph{et al}. (Cryptogr. Commun. 14(1), 145-159, 2022) when $\ell\neq \frac{e}{2}$.
\end{abstract}

\vspace{10pt}


\vspace{6pt}
}

\section{Introduction}\label{section1}

Let $q=p^{e}$ be a prime power and let $\ell$ be an integer with $0\leq\ell\leq e-1$.
Denote by $\mathrm{SLAut}(\mathbb{F}_{q}^{n})$ the group consisting of all semilinear isometries on $\mathbb{F}_{q}^{n}$.
Each element $\sigma\in\mathrm{SLAut}(\mathbb{F}_{q}^{n})$ can be expressed as $\sigma=(\tau,\pi_{s})$
which satisfies $\sigma(\mathbf{c})=\pi_{s}(\mathbf{c})M_{\tau}$ for any $\mathbf{c}\in\mathbb{F}_{q}^{n}$,
where the permutation $\tau$ in the symmetric group of order $n$ corresponds to an $n\times n$ monomial matrix $M_{\tau}$,
and $\pi_{s}$ denotes an automorphism of $\mathbb{F}_{q}$ with $1\leq s\leq e$ satisfying $\pi_{s}(a)=a^{p^{s}}$ for any $a\in\mathbb{F}_{q}$.
Recently, Carlet \emph{et al}. \cite{Carlet2019On} introduced the concept of $\sigma$ inner product for $\sigma\in\mathrm{SLAut}(\mathbb{F}_{q}^{n})$,
generalizing the one of Euclidean inner product, Hermitian inner product and $\ell$-Galois inner product.
From the concept of $\sigma$ inner product, it is clear that the $\sigma$ dual $\mathcal{C}^{\bot_{\sigma}}$ of a linear code $\mathcal{C}$
extends the Euclidean dual $\mathcal{C}^{\bot_{E}}$, Hermitian dual $\mathcal{C}^{\bot_{H}}$ and $\ell$-Galois dual $\mathcal{C}^{\bot_{\ell}}$ of it.

In recent years, some topics related to the intersection of two linear codes have received wide attention by many researchers, such as constructions of linear
complementary dual (LCD) codes
(see, e.g., \cite{Carlet2018Euclidean,Carlet2018Linear,Carlet2018Newcharacterization,Sok2018Constructions,Shi2021LCD,Massey1992}),
entanglement-assisted quantum error-correcting codes (EAQECCs) (see, e.g., \cite{Fang2020Euclidean,Luo2019MDS,Cao2021MDS,Sok2022On})
and asymmetric EAQECCs (see, e.g., \cite{Galindo2020Asymmetric,Huang2022Linear}).
We use the notations $Hull_{E}(\mathcal{C}):=\mathcal{C}\cap \mathcal{C}^{\perp_{E}}$, $Hull_{H}(\mathcal{C}):=\mathcal{C}\cap \mathcal{C}^{\perp_{H}}$
and $Hull_{\ell}(\mathcal{C}):=\mathcal{C}\cap \mathcal{C}^{\perp_{\ell}}$ to denote
Euclidean hull, Hermitian hull and $\ell$-Galois hull of a linear code $\mathcal{C}$, respectively.
In \cite{Guenda2018}, Guenda \emph{et al}. showed that for a linear code $\mathcal{C}$ with a generator matrix $G$ and a parity check matrix $H$,
the dimension of $Hull_{E}(\mathcal{C})$ \big(resp. $Hull_{H}(\mathcal{C})$\big) can be determined by
$\mathrm{rank}(GG^{T})$ \big(resp. $\mathrm{rank}(GG^{\dag})$\big) or $\mathrm{rank}(HH^{T})$ \big(resp. $\mathrm{rank}(HH^{\dag})$\big),
where $A^{\dag}:=\big[a_{j,i}^{\sqrt{q}}\big]$ for the matrix $A=[a_{i,j}]$ over $\mathbb{F}_{q}$.
For the dimension of $\ell$-Galois hull $Hull_{\ell}(\mathcal{C})$, Liu \emph{et al}. \cite{Liu2020New-EAQEC} proved that it can be determined by
$\mathrm{rank}(HH^{\ddag})$, where $A^{\ddag}:=[\pi_{e-\ell}(a_{j,i})]$ for the matrix $A=[a_{i,j}]$ over $\mathbb{F}_{q}$.
In \cite{Guenda2020Linear}, Guenda \emph{et al}. investigated a more general case on the intersection of two linear codes.
They proved that the dimension of $\mathcal{C}_{1}\cap\mathcal{C}_{2}$ can be determined by
$\mathrm{rank}(G_{i}H_{3-i}^{T})$, where $G_{i}$ (resp. $H_{i}$) is a generator matrix (resp. parity check matrix) of $\mathcal{C}_{i}$ for $i=1,2$.
They also showed that the dimension of $\mathcal{C}_{1}\cap\mathcal{C}_{2}$ can be determined by
$\mathrm{rank}\big(G_{i}\pi_{\ell}(H_{3-i}^{T})\big)$,
where $G_{i}$ (resp. $H_{i}$) is a generator matrix of $\mathcal{C}_{i}$ (resp. $\mathcal{C}_{i}^{\bot_{\ell}}$) for $i=1,2$.

In 2001, Blackmore and Norton \cite{Blackmore2001} proposed the concept of matrix-product codes over finite fields.
Matrix-product codes, labeled as $\mathcal{C}(A):=[\mathcal{C}_{1},\mathcal{C}_{2},\ldots,\mathcal{C}_{k}]\cdot A$,
are a class of linear codes with larger length by combining several constituent linear codes $\mathcal{C}_{1},\mathcal{C}_{2},\ldots,\mathcal{C}_{k}$
of the same length with an $k\times k$ defining matrix $A$.
In recent years, many interesting results on such codes were obtained, such as their minimum distances, duals, hulls and decoding algorithms.
It was shown in \cite{Blackmore2001,Hernando2009Construction,O2002Note} that the minimum distance (or lower bound of minimum distance)
of a matrix-product code is related to properties of the defining matrix and constituent codes.
The Euclidean dual and Hermitian dual of a matrix-product code were given by Blackmore and Norton in \cite{Blackmore2001} and by Zhang and Ge in \cite{Zhang2015Quantum}, respectively. In \cite{Liuhongwei2020Galois},
Liu and Pan presented the $\ell$-Galois hull of matrix-product codes over finite fields.
Besides, matrix-product codes over different rings were studied in \cite{Fan2014Matrix,Fan2014Homogeneous,Asch2008Matrix,Caoyuan2018Matrix}.
Notice that the decoding algorithms for matrix-product codes were provided in
\cite{Hernando2009Construction,Hernando2012List,Hernando2013Decoding}.
Matrix-product codes also play an important role in the fields of quantum error-correcting codes
(see, e.g., \cite{Liu2018On,Galindo2015New,Cao2020Constructioncaowang,Liu2019Entanglement-assisted})
and locally repairable codes (see, e.g., \cite{Luo2022Three}).

In this paper, we investigate general properties of linear codes associated with $\sigma$ duals for $\sigma\in\mathrm{SLAut}(\mathbb{F}_{q}^{n})$.
In the first part of this paper, we show that the dimension of the intersection $\mathcal{C}_{1}\cap\mathcal{C}_{2}$ of two linear codes
$\mathcal{C}_{1}$ and $\mathcal{C}_{2}$ can be determined by generator matrices of such codes and their
$\sigma$ duals $\mathcal{C}_{i}^{\bot_{\sigma}}$ for $i=1,2$,
generalizing the results for Euclidean case and $\ell$-Galois case considered in \cite{Guenda2020Linear}.
For $\mathcal{C}_{2}=\mathcal{C}_{1}^{\bot_{\sigma}}$, our result on the dimension of $\sigma$ hull of a linear code
generalizes the cases of Euclidean hull and Hermitian hull considered in \cite{Guenda2018},
and the case of $\ell$-Galois hull considered in \cite{Liu2020New-EAQEC} where the dimension is related to a parity check matrix of the linear code.
Our result reveals that the dimension of $\ell$-Galois hull of a linear code can be also associated with a generator matrix
of the linear code.

The second part of this paper is devoted to studying matrix-product codes.
We give a characterization on $\sigma$ duals of such codes, generalizing the cases of
Euclidean duals of them shown in \cite{Blackmore2001} and Hermitian duals of them shown in \cite{Zhang2015Quantum}, respectively.
Afterwards, the $\sigma$ hull of a matrix-product code is also presented,
which is a generalization of the case of $\ell$-Galois hull of a matrix-product code shown in \cite{Liuhongwei2020Galois}.
Notice that the case of Euclidean hull (resp. Hermitian hull) of a matrix-product code in our result actually provides a new manner for deriving a sufficient condition under which a matrix-product code is Euclidean self-orthogonal (resp. Hermitian self-orthogonal) or Euclidean dual-containing (resp. Hermitian dual-containing). We also investigate the intersection of a pair of matrix-product codes.

In the third part of this paper, we provide a necessary and sufficient condition under which any codeword of a generalized Reed-Solomon (GRS) code
or an extended GRS code is contained in its $\sigma$ dual,
generalizing the conditions for Euclidean case in \cite{Chen2018New}, Hermitian case in \cite{Fang2018Two} and $\ell$-Galois case in \cite{Cao2021MDS}.
As an application, we construct eleven families of $q$-ary MDS codes with new $\ell$-Galois hulls satisfying $2(e-\ell)\mid e$.
Note that the parameter $\ell$ of the $q$-ary MDS codes shown in the latest papers \cite{Cao2021MDS} and \cite{Fang2022New} satisfies $2\ell\mid e$.
Consequently, our MDS codes are not covered by \cite{Cao2021MDS} and \cite{Fang2022New} when $\ell\neq \frac{e}{2}$.

The rest of this paper is organized as follows.
Section \ref{section2} is the preliminaries of this paper.
In Section \ref{section3}, we characterize the dimension of the intersection of two linear codes and the dimension of $\sigma$ hull of a linear code.
In Section \ref{section4}, we explore $\sigma$ duals, $\sigma$ hulls and intersections of matrix-product codes.
In Section \ref{section5}, we construct eleven families of $q$-ary MDS codes with new $\ell$-Galois hulls, which are not covered by the latest literature.
Section \ref{section6} gives a conclusion of this paper.

\section{Preliminaries}\label{section2}

As usual, denote by $[n,k,d]_{q}$ a classical linear code over $\mathbb{F}_{q}$ with length $n$, dimension $k$ and minimum distance $d$.
The minimum distance $d$ of a linear code satisfies the well-known \emph{Singleton bound} $d\leq n-k+1$.
If $d=n-k+1$, then such a linear code is called a \emph{maximum distance separable (MDS) code}.

Following the notations introduced in \cite{Carlet2019On},
we call a mapping $\sigma: \mathbb{F}_{q}^{n}\rightarrow\mathbb{F}_{q}^{n}$ an \emph{isometry} if $d_{H}\big(\sigma(\mathbf{u}),\sigma(\mathbf{v})\big)$
$=d_{H}(\mathbf{u},\mathbf{v})$ 
for any $\mathbf{u},\mathbf{v}\in\mathbb{F}_{q}^{n}$,
where $d_{H}(\mathbf{u},\mathbf{v})$ is the Hamming distance of $\mathbf{u}$ and $\mathbf{v}$.
All isometries on $\mathbb{F}_{q}^{n}$ form a group, which is denoted by $\mathrm{Aut}(\mathbb{F}_{q}^{n})$.
Two codes $\mathcal{C}$ and $\mathcal{C}'$ are called \emph{isometric} if $\sigma(\mathcal{C})=\mathcal{C}'$ for some $\sigma\in\mathrm{Aut}(\mathbb{F}_{q}^{n})$.
If an isometry is linear, then it is called a linear isometry.
Let $\mathrm{MAut}(\mathbb{F}_{q}^{n})$ denote the \emph{monomial group} consisting of
the set of transforms with $n\times n$ monomial matrices over $\mathbb{F}_{q}$.
Notice that the group of all linear isometries of $\mathbb{F}_{q}^{n}$ corresponds to $\mathrm{MAut}(\mathbb{F}_{q}^{n})$.

For isometries that map subspaces onto subspaces, it is shown in \cite{Betten2006Error,Sendrier2013The}
that when $n\geq 3$, such isometries are exactly the semilinear mappings of the form
{\setlength\abovedisplayskip{0.15cm}
\setlength\belowdisplayskip{0.15cm}
\begin{align*}
\begin{split}
\sigma=(\tau,\pi):\ &\mathbb{F}_{q}^{n}\longrightarrow\mathbb{F}_{q}^{n}\\
&\ \ \mathbf{c}\longmapsto\tau\big(\pi(\mathbf{c})\big)
\end{split}
\end{align*}}with $\pi(\mathbf{c}):=\big(\pi(c_{1}),\pi(c_{2}),\ldots,\pi(c_{n})\big)$ for $\mathbf{c}=(c_{1},c_{2},\ldots,c_{n})\in\mathbb{F}_{q}^{n}$,
where $\tau$ is a linear isometry and $\pi$ is an automorphism of $\mathbb{F}_{q}$.
We denote by $\mathrm{SLAut}(\mathbb{F}_{q}^{n})$ the group of all semilinear isometries on $\mathbb{F}_{q}^{n}$.

\begin{remark}\label{remark2.1}
For any $\sigma=(\tau,\pi)\in \mathrm{SLAut}(\mathbb{F}_{q}^{n})$, the above definition implies that there exists an $n\times n$ monomial matrix $M_{\tau}$
over $\mathbb{F}_{q}$ which corresponds to $\tau$ such that
{\setlength\abovedisplayskip{0.15cm}
\setlength\belowdisplayskip{0.15cm}
\begin{equation}\label{eq2.1}
\sigma(\mathbf{c})=\tau\big(\pi(\mathbf{c})\big)=\pi(\mathbf{c})M_{\tau}
\end{equation}}for any $\mathbf{c}\in\mathbb{F}_{q}^{n}$.
Here, we can write the monomial matrix $M_{\tau}$ as $M_{\tau}=D_{\tau}P_{\tau}$, where $D_{\tau}$ is an invertible diagonal matrix,
and $P_{\tau}$ is a permutation matrix under $\tau$ such that the $\tau(i)$-th row of $P_{\tau}$ is exactly the $i$-th row of the identity matrix $I_{n}$
for $i=1,2,\ldots,n$. Consequently, it is clear that
{\setlength\abovedisplayskip{0.15cm}
\setlength\belowdisplayskip{0.15cm}
\begin{equation*}
(t_{1},t_{2},\ldots,t_{n})P_{\tau}=\big(t_{\tau(1)},t_{\tau(2)},\ldots,t_{\tau(n)}\big),
\end{equation*}
\begin{equation}\label{eq2.2}
(t_{1},t_{2},\ldots,t_{n})P_{\tau}^{T}=\big(t_{\tau^{-1}(1)},t_{\tau^{-1}(2)},\ldots,t_{\tau^{-1}(n)}\big).
\end{equation}}
\end{remark}

Carlet \emph{et al}. \cite{Carlet2019On} introduced the following useful concepts on $\sigma$ inner product and $\sigma$ dual of linear codes.

\begin{definition}\label{definition2.2}
(\!\!\cite{Carlet2019On})
Let $q=p^{e}$ be a prime power and let $\mathcal{C}$ be a linear code of length $n$ over $\mathbb{F}_{q}$. For $\sigma=(\tau,\pi)\in\mathrm{SLAut}(\mathbb{F}_{q}^{n})$, the \emph{$\sigma$ inner product} of
$\mathbf{a}\in\mathbb{F}_{q}^{n}$ and $\mathbf{b}\in\mathbb{F}_{q}^{n}$ is defined as
{\setlength\abovedisplayskip{0.15cm}
\setlength\belowdisplayskip{0.15cm}
\begin{equation*}
\langle\mathbf{a},\mathbf{b}\rangle_{\sigma}=\sum_{i=1}^{n}a_{i}c_{i},
\end{equation*}}where $\mathbf{a}=(a_{1},a_{2},\ldots,a_{n})$ and $\sigma(\mathbf{b})=(c_{1},c_{2},\ldots,c_{n})$.
The \emph{$\sigma$ dual} of $\mathcal{C}$ is defined as
{\setlength\abovedisplayskip{0.15cm}
\setlength\belowdisplayskip{0.15cm}
\begin{equation*}
\mathcal{C}^{\bot_{\sigma}}=\{\mathbf{a}\in\mathbb{F}_{q}^{n}|\ \langle\mathbf{a},\mathbf{b}\rangle_{\sigma}=0 \ \mathrm{for} \ \mathrm{all} \  \mathbf{b}\in\mathcal{C}\}.
\end{equation*}}
\end{definition}

With the notations above, denote $\sigma(\mathcal{C}):=\{\sigma(\mathbf{c})|\ \mathbf{c}\in\mathcal{C}\}$.
It is verified that $\mathcal{C}^{\bot_{\sigma}}=\big(\sigma(\mathcal{C})\big)^{\bot_{E}}$ (see \cite{Carlet2019On}).

\begin{remark}\label{remark2.3}
If $M_{\tau}=I_{n}$ and $\pi=\pi_{e-\ell}$ in Eq. (\ref{eq2.1}), where $\pi_{e-\ell}(a):=a^{p^{e-\ell}}$ with $0\leq \ell\leq e-1$
for each $a\in\mathbb{F}_{q}$,
then $\langle\mathbf{a},\mathbf{b}\rangle_{\sigma}=\langle\mathbf{a},\mathbf{b}\rangle_{e-\ell}:=\sum_{i=1}^{n}a_{i}b_{i}^{p^{e-\ell}}$,
which is the $(e-\ell)$-Galois inner product \cite{Fan2017Galois} of $\mathbf{a}$ and $\mathbf{b}$; $\mathcal{C}^{\bot_{\sigma}}=\mathcal{C}^{\bot_{\ell}}$,
which is the $\ell$-Galois dual \cite{Fan2017Galois} of $\mathcal{C}$. Further,

1) When $\ell=0$, then $\langle\mathbf{a},\mathbf{b}\rangle_{\sigma}=\langle\mathbf{a},\mathbf{b}\rangle_{E}:=\sum_{i=1}^{n}a_{i}b_{i}$,
which is the Euclidean inner product of $\mathbf{a}$ and $\mathbf{b}$; $\mathcal{C}^{\bot_{\sigma}}=\mathcal{C}^{\bot_{E}}$, which is the Euclidean dual of $\mathcal{C}$.

2) When $\ell=\frac{e}{2}$ for even $e$, then $\langle\mathbf{a},\mathbf{b}\rangle_{\sigma}=\langle\mathbf{a},\mathbf{b}\rangle_{H}:=\sum_{i=1}^{n}a_{i}b_{i}^{\sqrt{q}}$,
which is the Hermitian inner product of $\mathbf{a}$ and $\mathbf{b}$; $\mathcal{C}^{\bot_{\sigma}}=\mathcal{C}^{\bot_{H}}$, which is the Hermitian dual of $\mathcal{C}$.
\end{remark}

Let us fix the following notations in the rest of this paper.
\begin{itemize}
\setlength{\itemsep}{1pt}
\setlength{\parsep}{1pt}
\setlength{\parskip}{1pt}
\item A prime power $q$ is written as $q=p^{e}$, where $p$ is a prime number and $e$ is a positive integer.

\item $\mathbb{F}_{q}^{\ast}:=\mathbb{F}_{q}\backslash \{0\}$.

\item $\mathcal{M}(\mathbb{F}_{q},n\times m)$ denotes the set consisting of all $n\times m$ matrices over $\mathbb{F}_{q}$.
It is written as $\mathcal{M}(\mathbb{F}_{q},n)$ if $n=m$.

\item $A^{\ast}:=\big[a_{i,j}^{\sqrt{q}}\big]$ and $A^{\dag}:=\big[a_{j,i}^{\sqrt{q}}\big]$ for any matrix $A=[a_{i,j}]$ over $\mathbb{F}_{q}$.

\item $\mathcal{C}A:=\{\mathbf{c}A|\mathbf{c}\in\mathcal{C}\}$, where $\mathcal{C}$ is a linear code of length $n$ over $\mathbb{F}_{q}$ and
$A\in\mathcal{M}(\mathbb{F}_{q},n)$.

\item $\pi_{s}$ denotes the automorphism of $\mathbb{F}_{q}$  satisfying $\pi_{s}(a)=a^{p^{s}}$ for any $a\in\mathbb{F}_{q}$,
where $1\leq s\leq e$.

\item $\pi_{s}(A):=[\pi_{s}(a_{i,j})]$ for any matrix $A=[a_{i,j}]$ over $\mathbb{F}_{q}$.

\item  For $\sigma=(\tau,\pi_{s})\in\mathrm{SLAut}(\mathbb{F}_{q}^{n})$,
$\tau$ corresponds to a monomial matrix $M_{\tau}=D_{\tau}P_{\tau}$ with $D_{\tau}$ being an invertible diagonal matrix
and $P_{\tau}$ being a permutation matrix under $\tau$, that is to say, $\tau(\mathbf{u})=\mathbf{u}M_{\tau}$ for any $\mathbf{u}\in\mathbb{F}_{q}^{n}$.
Therefore, $\sigma(\mathbf{c})=\tau\big(\pi_{s}(\mathbf{c})\big)=\pi_{s}(\mathbf{c})M_{\tau}$ for any $\mathbf{c}\in\mathbb{F}_{q}^{n}$
(see also Eq. (\ref{eq2.1})).

\item $Hull_{\sigma}(\mathcal{C}):=\mathcal{C}\cap\mathcal{C}^{\bot_{\sigma}}$ is called the \emph{$\sigma$ hull} of the linear code $\mathcal{C}$,
where $\sigma\in \mathrm{SLAut}(\mathbb{F}_{q}^{n})$.

\item $Hull_{\ell}(\mathcal{C}):=\mathcal{C}\cap\mathcal{C}^{\bot_{\ell}}$ is called the \emph{$\ell$-Galois hull} of the linear code $\mathcal{C}$,
where $0\leq \ell\leq e-1$.

\item $Hull_{E}(\mathcal{C}):=\mathcal{C}\cap\mathcal{C}^{\bot_{E}}$ is called the \emph{Euclidean hull} of the linear code $\mathcal{C}$.

\item $Hull_{H}(\mathcal{C}):=\mathcal{C}\cap\mathcal{C}^{\bot_{H}}$ is called the \emph{Hermitian hull} of the linear code $\mathcal{C}$.

\end{itemize}

\section{The dimension of the intersection of two linear codes}\label{section3}

\begin{lemma}\label{lemma3.1}
Let $\mathcal{C}$ be an $[n,k]_{q}$ code with a parity check matrix $H$. Set $\sigma=(\tau,\pi_{s})\in\mathrm{SLAut}(\mathbb{F}_{q}^{n})$,
where $\tau$ corresponds to a monomial matrix $M_{\tau}=D_{\tau}P_{\tau}\in\mathcal{M}(\mathbb{F}_{q},n)$ and $1\leq s\leq e$. Then,
$\pi_{s}(H)(M_{\tau}^{-1})^{T}$ is a generator matrix of $\mathcal{C}^{\bot_{\sigma}}$.
\end{lemma}

\begin{proof}
Assume that $G$ generates $\mathcal{C}$. By Eq. (\ref{eq2.1}), we know $\pi_{s}(G)M_{\tau}$ generates $\sigma(\mathcal{C})$.
As
\begin{equation*}
\pi_{s}(G)M_{\tau}\big(\pi_{s}(H)(M_{\tau}^{-1})^{T}\big)^{T}=\pi_{s}(GH^{T})=0,
\end{equation*}
$\mathcal{C}^{\bot_{\sigma}}=\big(\sigma(\mathcal{C})\big)^{\bot_{E}}$ and $\pi$ preserves the dimension,
then $\pi_{s}(H)(M_{\tau}^{-1})^{T}$ generates $\mathcal{C}^{\bot_{\sigma}}$. $\hfill\square$
\end{proof}

\begin{lemma}\label{lemma3.2}
Let $\mathcal{C}$ be an $[n,k]_{q}$ code. Set $\sigma=(\tau,\pi_{s})\in\mathrm{SLAut}(\mathbb{F}_{q}^{n})$,
where $\tau$ corresponds to a monomial matrix $M_{\tau}=D_{\tau}P_{\tau}\in\mathcal{M}(\mathbb{F}_{q},n)$ and $1\leq s\leq e$. Then,
$\sigma(\mathcal{C}^{\bot_{E}})=\big(\sigma(\mathcal{C})\big)^{\bot_{E}}M_{\tau}^{T}M_{\tau}$.
\end{lemma}

\begin{proof}
Assume that $H$ generates $\mathcal{C}^{\bot_{E}}$. Then, by Eq. (\ref{eq2.1}), $\pi_{s}(H)M_{\tau}$ generates $\sigma(\mathcal{C}^{\bot_{E}})$.
By Lemma \ref{lemma3.1}, $\pi_{s}(H)(M_{\tau}^{-1})^{T}$ generates $\big(\sigma(\mathcal{C})\big)^{\bot_{E}}$.
Thus, we obtain $\sigma(\mathcal{C}^{\bot_{E}})=\big(\sigma(\mathcal{C})\big)^{\bot_{E}}M_{\tau}^{T}M_{\tau}$. $\hfill\square$
\end{proof}

\begin{lemma}\label{lemma3.3}
Let $\mathcal{C}$ be an $[n,k]_{q}$ code. Set $\sigma=(\tau,\pi_{s})\in\mathrm{SLAut}(\mathbb{F}_{q}^{n})$,
where $\tau$ corresponds to a monomial matrix $M_{\tau}=D_{\tau}P_{\tau}\in\mathcal{M}(\mathbb{F}_{q},n)$ and $1\leq s\leq e$.
If $H$ is a generator matrix of $\mathcal{C}^{\bot_{\sigma}}$, then $\pi_{e-s}(HM_{\tau}^{T})$ is a generator matrix of $\mathcal{C}^{\bot_{E}}$.
\end{lemma}

\begin{proof}
Note that
{\setlength\abovedisplayskip{0.15cm}
\setlength\belowdisplayskip{0.15cm}
\begin{align*}
\mathcal{C}^{\bot_{\sigma}}=\big(\sigma(\mathcal{C})\big)^{\bot_{E}}=\sigma(\mathcal{C}^{\bot_{E}})(M_{\tau}^{T}M_{\tau})^{-1}
=\big(\pi_{s}(\mathcal{C}^{\bot_{E}})M_{\tau}\big)(M_{\tau}^{T}M_{\tau})^{-1}=\pi_{s}(\mathcal{C}^{\bot_{E}})(M_{\tau}^{T})^{-1},
\end{align*}}where the second equality follows from Lemma \ref{lemma3.2}.
Since $M_{\tau}$ is invertible, we know that $HM_{\tau}^{T}$ is a generator matrix of $\mathcal{C}^{\bot_{\sigma}}M_{\tau}^{T}$.
That is, $HM_{\tau}^{T}$ is a generator matrix of $\pi_{s}(\mathcal{C}^{\bot_{E}})$,
which implies that $\pi_{e-s}(HM_{\tau}^{T})$ is a generator matrix of $\mathcal{C}^{\bot_{E}}$
since $\pi_{e-s}\big(\pi_{s}(\mathcal{C}^{\bot_{E}})\big)=\mathcal{C}^{\bot_{E}}$. $\hfill\square$
\end{proof}

\vspace{6pt}

In the following theorem, we establish a relationship between the dimension of the intersection of two linear codes and generator matrices of such codes
and their $\sigma$ duals.

\begin{theorem}\label{theorem3.4}
Let $\mathcal{C}_{i}$ be an $[n,k_{i}]_{q}$ code with a generator matrix $G_{i}$ for $i=1,2$. Set $\sigma=(\tau,\pi_{s})\in\mathrm{SLAut}(\mathbb{F}_{q}^{n})$, where $\tau$ corresponds to a monomial matrix $M_{\tau}=D_{\tau}P_{\tau}\in\mathcal{M}(\mathbb{F}_{q},n)$ and $1\leq s\leq e$.
Let $H_{i}$ be a generator matrix of $\mathcal{C}_{i}^{\bot_{\sigma}}$ for $i=1,2$.
Suppose $\mathrm{dim}_{\mathbb{F}_{q}}(\mathcal{C}_{1}\cap\mathcal{C}_{2})=\Delta$, then

\vspace{4pt}

(1) $\mathrm{rank}\big(\pi_{e-s}(H_{2}M_{\tau}^{T})G_{1}^{T}\big)=\mathrm{rank}\big(G_{1}\pi_{e-s}(M_{\tau}H_{2}^{T})\big)=k_{1}-\Delta$.

\vspace{4pt}

(2) $\mathrm{rank}\big(\pi_{e-s}(H_{1}M_{\tau}^{T})G_{2}^{T}\big)=\mathrm{rank}\big(G_{2}\pi_{e-s}(M_{\tau}H_{1}^{T})\big)=k_{2}-\Delta$.
\end{theorem}

\begin{proof}
(1) As $G_{1}\pi_{e-s}(M_{\tau}H_{2}^{T})=\big(\pi_{e-s}(H_{2}M_{\tau}^{T})G_{1}^{T}\big)^{T}$,
we only need to show $\mathrm{rank}\big(\pi_{e-s}(H_{2}M_{\tau}^{T})G_{1}^{T}\big)=k_{1}-\Delta$.
Note that $n\geq\mathrm{dim}_{\mathbb{F}_{q}}(\mathcal{C}_{1}+\mathcal{C}_{2})=k_{1}+k_{2}-\Delta$,
we obtain $n-k_{3-i}\geq k_{i}-\Delta$ for $i=1,2$. Let us consider the following two cases.

{\bfseries Case (i)}: When $\Delta=k_{1}$, we have $\mathcal{C}_{1}\subseteq\mathcal{C}_{2}$.
It follows from Lemma \ref{lemma3.3} that $\pi_{e-s}(H_{2}M_{\tau}^{T})$ is a generator matrix of $\mathcal{C}_{2}^{\bot_{E}}$.
Thus, we have $G_{1}\big(\pi_{e-s}(H_{2}M_{\tau}^{T})\big)^{T}=0$, i.e., $\pi_{e-s}(H_{2}M_{\tau}^{T})G_{1}^{T}=0$.
Hence, the desired result holds.

{\bfseries Case (ii)}: When $\Delta<k_{1}$, we may assume that $\{\mathbf{g}_{1},\mathbf{g}_{2},\ldots,\mathbf{g}_{\Delta}\}$
is a basis of $\mathcal{C}_{1}\cap\mathcal{C}_{2}$. It can be extended to be a basis, labeled as
$\{\mathbf{g}_{1},\mathbf{g}_{2},\ldots,\mathbf{g}_{\Delta},\mathbf{g}_{\Delta+1},\ldots,\mathbf{g}_{k_{1}}\}$,
of $\mathcal{C}_{1}$. Then
\begin{align*}
\widetilde{G_{1}}:=\left[
\begin{array}{c}
\mathbf{g}_{1}\\[0.01pt]
\vdots\\[0.01pt]
\mathbf{g}_{\Delta}\\\hline
\mathbf{g}_{\Delta+1}\\[0.01pt]
\vdots\\[0.01pt]
\mathbf{g}_{k_{1}}\\
\end{array}\right]
=\left[
\begin{array}{c}
\mathbf{g}_{1}\\[0.01pt]
\vdots\\[0.01pt]
\mathbf{g}_{\Delta}\\\hline
R\\
\end{array}\right] \ \left(\mathrm{with}\ R:=\left[
\begin{array}{c}
\mathbf{g}_{\Delta+1}\\[0.01pt]
\vdots\\[0.01pt]
\mathbf{g}_{k_{1}}\\
\end{array}\right]\right)
\end{align*}
is a generator matrix of $\mathcal{C}_{1}$.
It follows that there exists an invertible matrix $F$ such that $G_{1}=F\widetilde{G_{1}}$. Thus,
$\mathrm{rank}\big(\pi_{e-s}(H_{2}M_{\tau}^{T})G_{1}^{T}\big)=\mathrm{rank}\Big(\pi_{e-s}(H_{2}M_{\tau}^{T})\widetilde{G_{1}}^{T}F^{T}\Big)
=\mathrm{rank}\Big(\pi_{e-s}(H_{2}M_{\tau}^{T})\widetilde{G_{1}}^{T}\Big)$.

Since $\mathbf{g}_{1},\ldots,\mathbf{g}_{\Delta}\in\mathcal{C}_{2}$, it follows from Lemma \ref{lemma3.3} that
\begin{align*}
\pi_{e-s}(H_{2}M_{\tau}^{T})\widetilde{G_{1}}^{T}
&=\pi_{e-s}(H_{2}M_{\tau}^{T})\big[\mathbf{g}_{1}^{T},\ldots,\mathbf{g}_{\Delta}^{T},R^{T}\big]\\
&=\big[\mathbf{0},\ldots,\mathbf{0},\pi_{e-s}(H_{2}M_{\tau}^{T})R^{T}\big].
\end{align*}

As $\pi_{e-s}(H_{2}M_{\tau}^{T})R^{T}$ is an $(n-k_{2})\times(k_{1}-\Delta)$ matrix, we have
$\mathrm{rank}\big(\pi_{e-s}(H_{2}M_{\tau}^{T})R^{T}\big)\leq k_{1}-\Delta$.
Let us now prove that $\mathrm{rank}\big(\pi_{e-s}(H_{2}M_{\tau}^{T})R^{T}\big)=k_{1}-\Delta$.
Suppose
$\mathrm{rank}\big(\pi_{e-s}(H_{2}M_{\tau}^{T})R^{T}\big)<k_{1}-\Delta$,
then there exists a nonzero vector $\mathbf{u}\in\mathbb{F}_{q}^{k_{1}-\Delta}$ such that
$\pi_{e-s}(H_{2}M_{\tau}^{T})R^{T}\mathbf{u}^{T}=\mathbf{0}$,
which implies that $\mathbf{u}R\in\mathcal{C}_{2}$ by Lemma \ref{lemma3.3}. Further, we obtain $\mathbf{u}R\neq\mathbf{0}$,
otherwise, $\mathbf{g}_{\Delta+1},\ldots,\mathbf{g}_{k_{1}}$ is linearly dependent and so is
$\mathbf{g}_{1},\ldots,\mathbf{g}_{k_{1}}$, which turns to be a contradiction. So $\mathbf{u}R\in\mathcal{C}_{2}\backslash\{\mathbf{0}\}$.

Due to the linearity of $\mathcal{C}_{1}$, it is obvious that $\mathbf{u}R\in\mathcal{C}_{1}$. Thus
$\mathbf{u}R\in\mathcal{C}_{1}\cap\mathcal{C}_{2}=\mathrm{span}\{\mathbf{g}_{1},\mathbf{g}_{2},\ldots,\mathbf{g}_{\Delta}\}$.
Denote $\mathbf{u}=(u_{1},u_{2},\ldots,u_{k_{1}-\Delta})$, then there exists $v_{i}\in\mathbb{F}_{q}$, $i=1,2,\ldots,\Delta$, such that
\begin{equation*}
\sum_{i=1}^{k_{1}-\Delta}u_{i}\mathbf{g}_{\Delta+i}=\sum_{i=1}^{\Delta}v_{i}\mathbf{g}_{i}.
\end{equation*}
Since $\mathbf{g}_{1},\ldots,\mathbf{g}_{\Delta},\mathbf{g}_{\Delta+1},\ldots,\mathbf{g}_{k_{1}}$ is linearly independent, we immediately obtain that
{\setlength\abovedisplayskip{0.15cm}
\setlength\belowdisplayskip{0.15cm}
\begin{equation*}
u_{1}=\cdots=u_{k_{1}-\Delta}=v_{1}=\cdots=v_{\Delta}=0.
\end{equation*}}Then $\mathbf{u}=\mathbf{0}$, which yields a contradiction.
Hence, we obtain $\mathrm{rank}\big(\pi_{e-s}(H_{2}M_{\tau}^{T})R^{T}\big)=k_{1}-\Delta$
and thus $\mathrm{rank}\big(\pi_{e-s}(H_{2}M_{\tau}^{T})G_{1}^{T}\big)=k_{1}-\Delta$. So the desired result holds.

\vspace{4pt}

(2) It is easily achieved by interchanging $\mathcal{C}_{1}$ and $\mathcal{C}_{2}$ in part (1). $\hfill\square$
\end{proof}

\begin{remark} In Theorem \ref{theorem3.4}: i) if $s=e-\ell$ with $0\leq \ell\leq e-1$, and $M_{\tau}=I_{n}$, then
we get the result for $\ell$-Galois case in \cite[Remark 2.1]{Guenda2020Linear};
ii) if $\ell=0$ and $M_{\tau}=I_{n}$, then we get the result for Euclidean case in \cite[Theorem 2.1]{Guenda2020Linear};
iii) if $\ell=\frac{e}{2}$ for even $e$, and $M_{\tau}=I_{n}$, then we get the result for Hermitian case.
\end{remark}

Next, let us give a characterization on the dimension of $\sigma$ hull of a linear code.

\begin{theorem}\label{theorem3.6}
Let $\mathcal{C}$ be an $[n,k]_{q}$ code with a generator matrix $G$. Set $\sigma=(\tau,\pi_{s})\in\mathrm{SLAut}(\mathbb{F}_{q}^{n})$,
where $\tau$ corresponds to a monomial matrix $M_{\tau}=D_{\tau}P_{\tau}\in\mathcal{M}(\mathbb{F}_{q},n)$ and $1\leq s\leq e$.
Let $H$ be a generator matrix of $\mathcal{C}^{\bot_{\sigma}}$.
Suppose $\mathrm{dim}_{\mathbb{F}_{q}}\big(Hull_{\sigma}(\mathcal{C})\big)=\Delta$, then

\vspace{4pt}

(1) $\mathrm{rank}\big(G(\pi_{s}(G)M_{\tau})^{T}\big)=\mathrm{rank}(\pi_{s}(G)M_{\tau}G^{T})=k-\Delta$.

\vspace{4pt}

(2) $\mathrm{rank}\big(\pi_{e-s}(HM_{\tau}^{T})H^{T}\big)=\mathrm{rank}\big(H\pi_{e-s}(M_{\tau}H^{T})\big)=n-k-\Delta$.
\end{theorem}

\begin{proof}
(1) Consider $\mathcal{C}_{2}=\mathcal{C}_{1}^{\bot_{\sigma}}$ in Theorem \ref{theorem3.4}. First, we are ready to show that
$\mathrm{rank}\big(G_{1}\pi_{e-s}(M_{\tau}H_{2}^{T})\big)=\mathrm{rank}\big(G_{1}(\pi_{s}(G_{1})M_{\tau})^{T}\big)$,
i.e., $\mathrm{rank}\Big(G_{1}\big(\pi_{e-s}(H_{2}M_{\tau}^{T})\big)^{T}\Big)=\mathrm{rank}\big(G_{1}(\pi_{s}(G_{1})M_{\tau})^{T}\big)$.
Note that $\pi_{e-s}(H_{2}M_{\tau}^{T})$ is a parity check matrix of $\mathcal{C}_{2}$ by Lemma \ref{lemma3.3}.
Besides, by $\mathcal{C}_{2}=\mathcal{C}_{1}^{\bot_{\sigma}}=(\pi_{s}(\mathcal{C}_{1})M_{\tau})^{\bot_{E}}$ we know
that $\pi_{s}(G_{1})M_{\tau}$ is also a parity check matrix of $\mathcal{C}_{2}$.
Then, there exists an invertible matrix $\widetilde{F}$ such that $\pi_{e-s}(H_{2}M_{\tau}^{T})=\widetilde{F}\pi_{s}(G_{1})M_{\tau}$
and thus
\begin{equation*}
\mathrm{rank}\Big(G_{1}\big(\pi_{e-s}(H_{2}M_{\tau}^{T})\big)^{T}\Big)
=\mathrm{rank}\Big(G_{1}\big(\pi_{s}(G_{1})M_{\tau}\big)^{T}\widetilde{F}^{T}\Big)
=\mathrm{rank}\big(G_{1}(\pi_{s}(G_{1})M_{\tau})^{T}\big).
\end{equation*}
That is, $\mathrm{rank}\big(G_{1}\pi_{e-s}(M_{\tau}H_{2}^{T})\big)=\mathrm{rank}\big(G_{1}(\pi_{s}(G_{1})M_{\tau})^{T}\big)$.

Further, take $\mathcal{C}_{1}=\mathcal{C}$ in Theorem \ref{theorem3.4}, then $\mathrm{rank}\big(G(\pi_{s}(G)M_{\tau})^{T}\big)=k-\Delta$.
As $G(\pi_{s}(G)M_{\tau})^{T}=(\pi_{s}(G)M_{\tau}G^{T})^{T}$, we get $\mathrm{rank}(\pi_{s}(G)M_{\tau}G^{T})=k-\Delta$.
This completes part (1).

\vspace{6pt}

(2) For any $\mathbf{c}\in Hull_{\sigma}(\mathcal{C})$, since $\mathbf{c}\in\mathcal{C}^{\bot_{\sigma}}$, there exists a vector
$\mathbf{f}=(f_{1},f_{2},\ldots,f_{n-k})\in\mathbb{F}_{q}^{n-k}$ such that $\mathbf{c}=\mathbf{f}H\in\mathcal{C}$.
By Lemma \ref{lemma3.3}, $\pi_{e-s}(HM_{\tau}^{T})$ is a parity check matrix of $\mathcal{C}$, which implies that
$\pi_{e-s}(HM_{\tau}^{T})\mathbf{c}^{T}=\mathbf{0}$, i.e., $\pi_{e-s}(HM_{\tau}^{T})H^{T}\mathbf{f}^{T}=\mathbf{0}$.
So $\mathrm{dim}_{\mathbb{F}_{q}}\big(Hull_{\sigma}(\mathcal{C})\big)=n-k-\mathrm{rank}\big(\pi_{e-s}(HM_{\tau}^{T})H^{T}\big)$,
i.e., $\mathrm{rank}\big(\pi_{e-s}(HM_{\tau}^{T})H^{T}\big)=n-k-\Delta$.
Note that $\big(\pi_{e-s}(HM_{\tau}^{T})H^{T}\big)^{T}=H\pi_{e-s}(M_{\tau}H^{T})$, we have $\mathrm{rank}\big(H\pi_{e-s}(M_{\tau}H^{T})\big)=n-k-\Delta$.
This completes part (2). $\hfill\square$
\end{proof}

\vspace{6pt}

Take $s=e-\ell$ with $0\leq \ell\leq e-1$, and $M_{\tau}=I_{n}$ in Theorem \ref{theorem3.6}, then $Hull_{\sigma}(\mathcal{C})=Hull_{\ell}(\mathcal{C})$ and thus
the following corollary for $\ell$-Galois case is immediately obtained.

\begin{corollary}\label{corollary3.7}
Let $\mathcal{C}$ be an $[n,k]_{q}$ code with a generator matrix $G$.
Let $H$ be a generator matrix of $\mathcal{C}^{\bot_{\ell}}$.
Suppose $\mathrm{dim}_{\mathbb{F}_{q}}\big(Hull_{\ell}(\mathcal{C})\big)=\Delta$, then

\vspace{4pt}

(1) $\mathrm{rank}\big(G\pi_{e-\ell}(G^{T})\big)=\mathrm{rank}\big(\pi_{e-\ell}(G)G^{T}\big)=k-\Delta$.

\vspace{4pt}

(2) $\mathrm{rank}\big(\pi_{\ell}(H)H^{T}\big)=\mathrm{rank}\big(H\pi_{\ell}(H^{T})\big)=n-k-\Delta$.

\end{corollary}

\begin{remark}\label{remark3.8}
With the notations in Corollary \ref{corollary3.7}.
When $H$ is a generator matrix of $\mathcal{C}^{\bot_{\ell}}$, it is not difficult to verify that
$\pi_{\ell}(H)$ is a parity check matrix of $\mathcal{C}$.
Suppose $\widehat{H}$ is a parity check matrix of $\mathcal{C}$, then there exists an invertible matrix $W$ such that
$\widehat{H}=W\pi_{\ell}(H)$. Hence $\pi_{e-\ell}(\widehat{H}^{T})=H^{T}\pi_{e-\ell}(W^{T})$.
It follows from Corollary \ref{corollary3.7} (2) that
\begin{equation*}
\mathrm{rank}\Big(\widehat{H}\pi_{e-\ell}\big(\widehat{H}^{T}\big)\Big)=\mathrm{rank}\Big(W\pi_{\ell}(H)H^{T}\pi_{e-\ell}\big(W^{T}\big)\Big)
=\mathrm{rank}\big(\pi_{\ell}(H)H^{T}\big)=n-k-\Delta,
\end{equation*}
which is exactly the result shown in \cite[Lemma 3.1]{Liu2020New-EAQEC}.
\end{remark}

\begin{remark}\label{remark3.9}
Taking $\ell=0$ and $\ell=\frac{e}{2}$ for even $e$, respectively, in Corollary \ref{corollary3.7}, then we get the results for Euclidean case
and Hermitian case provided in Propositions 3.1 and 3.2 of \cite{Guenda2018}, respectively.
\end{remark}

\section{$\sigma$ duals, $\sigma$ hulls and intersections of matrix-product codes}\label{section4}
In 2001, Blackmore and Norton \cite{Blackmore2001} proposed a new kind of long linear codes called matrix-product codes that are made up of several commensurate shorter linear codes with a defining matrix. In this section, we will explore the $\sigma$ duals, $\sigma$ hulls and intersections of certain kinds of matrix-product codes. Some of the results provided in this section are the generalizations of those in
\cite{Blackmore2001,Zhang2015Quantum,Liuhongwei2020Galois,Jitman2017,Mankean2016Matrixproduct,Liu2018On}.

\subsection{Basics of matrix-product codes}\label{subsection2.3}

\begin{definition}\label{definition4.1}
(\!\!\cite{Blackmore2001}) Let $\mathcal{C}_{1},\mathcal{C}_{2},\ldots,\mathcal{C}_{k}$ be linear codes of the same length $n$ over $\mathbb{F}_{q}$.
Let $A=[a_{i,j}]_{\mathop{}_{1\leq j\leq t}^{1\leq i\leq k}}\in \mathcal{M}(\mathbb{F}_{q},k\times t)$ be full rank with $k\leq t$.
A \emph{matrix-product code}
$\mathcal{C}(A):=[\mathcal{C}_{1},\mathcal{C}_{2},\ldots,\mathcal{C}_{k}]\cdot A$
with a \emph{defining matrix} $A$
is defined as the set of all matrix-products $[\mathbf{c}_{1},\mathbf{c}_{2},\ldots,\mathbf{c}_{k}]\cdot A$,
where $\mathbf{c}_{i}=(c_{1,i},c_{2,i},\ldots,c_{n,i})^{T}\in\mathcal{C}_{i}$ for $i=1,2,\ldots,k$.
A typical codeword $\mathbf{c}=[\mathbf{c}_{1},\mathbf{c}_{2},\ldots,\mathbf{c}_{k}]\cdot A$ of $\mathcal{C}(A)$ is an
$n\times t$ matrix:
\begin{equation*}
\mathbf{c}=\left[
\begin{array}{cccc}
\sum_{i=1}^k c_{1,i}a_{i,1}& \sum_{i=1}^k c_{1,i}a_{i,2}& \cdots &\sum_{i=1}^k c_{1,i}a_{i,t}\\
\sum_{i=1}^k c_{2,i}a_{i,1}& \sum_{i=1}^k c_{2,i}a_{i,2}& \cdots &\sum_{i=1}^k c_{2,i}a_{i,t}\\
\vdots&\vdots&\ddots&\vdots\\
\sum_{i=1}^k c_{n,i}a_{i,1}& \sum_{i=1}^k c_{n,i}a_{i,2}& \cdots &\sum_{i=1}^k c_{n,i}a_{i,t}\\
\end{array}\right].
\end{equation*}
\end{definition}
Reading the entries of the $n\times t$ matrix above in column-major order, any codeword of $\mathcal{C}(A)$ can be viewed as a row vector of length $tn$, i.e.,
$\mathbf{c}=\big[\sum_{i=1}^k a_{i,1}\mathbf{c}_{i},\sum_{i=1}^k a_{i,2}\mathbf{c}_{i},\ldots,\sum_{i=1}^k a_{i,t}\mathbf{c}_{i}\big]$,
where $\mathbf{c}_{i}=(c_{1,i},c_{2,i},\ldots,c_{n,i})\in\mathcal{C}_{i}$ is regarded as an $1\times n$ row vector for $i=1,2,\ldots,k$.

\begin{remark}\label{remark4.2}
In \cite{Liuhongwei2020Galois}, Liu and Pan showed that the codeword $[\mathbf{c}_{1},\mathbf{c}_{2},\ldots,\mathbf{c}_{k}]\cdot A$ of
$[\mathcal{C}_{1},\mathcal{C}_{2},\ldots,\mathcal{C}_{k}]\cdot A$ for column vectors $\mathbf{c}_{1},\mathbf{c}_{2},\ldots,\mathbf{c}_{k}\in \mathcal{C}_{i}$
is exactly the codeword $[\mathbf{c}_{1},\mathbf{c}_{2},\ldots,\mathbf{c}_{k}](A\otimes I_{n})$ of
$[\mathcal{C}_{1},\mathcal{C}_{2},\ldots,\mathcal{C}_{k}](A\otimes I_{n})$ for row vectors
$\mathbf{c}_{1},\mathbf{c}_{2},\ldots,\mathbf{c}_{k}\in \mathcal{C}_{i}$, where the multiplication in the latter is the same as usual.
\end{remark}

The following lemma characterizes the Euclidean dual of a matrix-product code.

\begin{lemma}\label{lemma4.3}
(\!\!\cite[Proposition 6.2]{Blackmore2001})
Let $\mathcal{C}(A)=[\mathcal{C}_{1},\mathcal{C}_{2},\ldots,\mathcal{C}_{k}]\cdot A$,
where $\mathcal{C}_{i}$ is an $[n,m_{i}]_{q}$ code for $i=1,2,\ldots,k$ and $A\in\mathcal{M}(\mathbb{F}_{q},k)$ is invertible.
Then the Euclidean dual of $\mathcal{C}(A)$ is
\begin{equation*}
\mathcal{C}(A)^{\bot_{E}}=\big[\mathcal{C}_{1}^{\bot_{E}},\mathcal{C}_{2}^{\bot_{E}},\ldots,\mathcal{C}_{k}^{\bot_{E}}\big]\cdot (A^{-1})^{T}.
\end{equation*}
\end{lemma}

\subsection{$\sigma$ duals of matrix-product codes}

Let us give an expression on $\sigma$ duals of matrix-product codes over finite fields in the following theorem.

\begin{theorem}\label{theorem4.4}
Let $\mathcal{C}(A)=[\mathcal{C}_{1},\mathcal{C}_{2},\ldots,\mathcal{C}_{k}]\cdot A$,
where $\mathcal{C}_{i}$ is an $[n,m_{i}]_{q}$ code for $i=1,2,\ldots,k$ and $A\in\mathcal{M}(\mathbb{F}_{q},k)$ is invertible.
Set $\sigma=(\tau,\pi_{s})\in\mathrm{SLAut}(\mathbb{F}_{q}^{kn})$ and $\sigma_{i}=(\tau_{i},\pi_{s})\in\mathrm{SLAut}(\mathbb{F}_{q}^{n})$,
where $\tau$ corresponds to a monomial matrix $M_{\tau}=D_{\tau}P_{\tau}\in\mathcal{M}(\mathbb{F}_{q},kn)$,
$\tau_{i}$ corresponds to a monomial matrix $M_{\tau_{i}}=D_{\tau_{i}}P_{\tau_{i}}\in\mathcal{M}(\mathbb{F}_{q},n)$ for $i=1,2\ldots,k$, and $1\leq s\leq e$.
Suppose $B$ satisfies
\begin{equation}\label{eq4.1}
(B\otimes I_{n})M_{\tau}^{T}\big(\pi_{s}(A)^{T}\otimes I_{n}\big)
=\mathrm{diag}\big(M_{\tau_{1}}^{T},M_{\tau_{2}}^{T},\ldots,M_{\tau_{k}}^{T}\big).
\end{equation}
Then the $\sigma$ dual of $\mathcal{C}(A)$ is
\begin{equation*}
\mathcal{C}(A)^{\bot_{\sigma}}=\big[\mathcal{C}_{1}^{\bot_{\sigma_{1}}},\mathcal{C}_{2}^{\bot_{\sigma_{2}}},\dots,\mathcal{C}_{k}^{\bot_{\sigma_{k}}}\big]\cdot B.
\end{equation*}
\end{theorem}

\begin{proof}
On the one hand,
\begin{align*}
\mathcal{C}(A)^{\bot_{\sigma}}
&=\sigma\big(\mathcal{C}(A)\big)^{\bot_{E}}\\
&=\sigma\big(\mathcal{C}(A)^{\bot_{E}}\big)M_{\tau}^{-1}(M_{\tau}^{T})^{-1}\\
&=\sigma\Big(\big[\mathcal{C}_{1}^{\bot_{E}},\mathcal{C}_{2}^{\bot_{E}},\ldots,\mathcal{C}_{k}^{\bot_{E}}\big]\big((A^{-1})^{T}\otimes I_{n}\big)\Big)
M_{\tau}^{-1}(M_{\tau}^{T})^{-1}\\
&=\pi_{s}\Big(\big[\mathcal{C}_{1}^{\bot_{E}},\mathcal{C}_{2}^{\bot_{E}},\ldots,\mathcal{C}_{k}^{\bot_{E}}\big]
\big((A^{-1})^{T}\otimes I_{n}\big)\Big)M_{\tau}M_{\tau}^{-1}(M_{\tau}^{T})^{-1}\\
&=\big[\pi_{s}(\mathcal{C}_{1}^{\bot_{E}}),\pi_{s}(\mathcal{C}_{2}^{\bot_{E}}),\ldots,\pi_{s}(\mathcal{C}_{k}^{\bot_{E}})\big]
\Big(\pi_{s}(A^{-1})^{T}\otimes I_{n}\Big)(M_{\tau}^{T})^{-1},
\end{align*}
where the second equality follows from Lemma \ref{lemma3.2} while the third equality follows from Remark \ref{remark4.2} and Lemma \ref{lemma4.3}.

On the other hand,
\begin{align*}
\big[\mathcal{C}_{1}^{\bot_{\sigma_{1}}},\mathcal{C}_{2}^{\bot_{\sigma_{2}}},\dots,\mathcal{C}_{k}^{\bot_{\sigma_{k}}}\big]\cdot B
&=\Big[\big(\sigma_{1}(\mathcal{C}_{1})\big)^{\bot_{E}},\big(\sigma_{2}(\mathcal{C}_{2})\big)^{\bot_{E}},\ldots,
\big(\sigma_{k}(\mathcal{C}_{k})\big)^{\bot_{E}}\Big]\cdot B\\
&=\Big[\sigma_{1}(\mathcal{C}_{1}^{\bot_{E}})M_{\tau_{1}}^{-1}(M_{\tau_{1}}^{T})^{-1},
\ldots,\sigma_{k}(\mathcal{C}_{k}^{\bot_{E}})M_{\tau_{k}}^{-1}(M_{\tau_{k}}^{T})^{-1}\Big]\cdot B\\
&=\Big[\pi_{s}(\mathcal{C}_{1}^{\bot_{E}})(M_{\tau_{1}}^{T})^{-1},
\ldots,\pi_{s}(\mathcal{C}_{k}^{\bot_{E}})(M_{\tau_{k}}^{T})^{-1}\Big]\cdot B\\
&=\Big[\pi_{s}(\mathcal{C}_{1}^{\bot_{E}}),\ldots,\pi_{s}(\mathcal{C}_{k}^{\bot_{E}})\Big]
\mathrm{diag}\big((M_{\tau_{1}}^{T})^{-1},\ldots,(M_{\tau_{k}}^{T})^{-1}\big)(B\otimes I_{n}),
\end{align*}
where the second and the fourth equality follows from Lemma \ref{lemma3.2} and Remark \ref{remark4.2}, respectively.

By Eq. (\ref{eq4.1}), it is easy to know that
\begin{equation*}
\Big(\pi_{s}(A^{-1})^{T}\otimes I_{n}\Big)(M_{\tau}^{T})^{-1}=\mathrm{diag}\big((M_{\tau_{1}}^{T})^{-1},\ldots,(M_{\tau_{k}}^{T})^{-1}\big)(B\otimes I_{n}),
\end{equation*}
which implies that $\mathcal{C}(A)^{\bot_{\sigma}}=\big[\mathcal{C}_{1}^{\bot_{\sigma_{1}}},\mathcal{C}_{2}^{\bot_{\sigma_{2}}},\dots,\mathcal{C}_{k}^{\bot_{\sigma_{k}}}\big]
\cdot B$. $\hfill\square$
\end{proof}

\begin{remark}\label{remark4.5}
1) In Theorem \ref{theorem4.4}, if $\tau_{1}=\ldots=\tau_{k}=\widetilde{\tau}$ for some $\widetilde{\tau}$, then the $\sigma$ dual of $\mathcal{C}(A)$ is
\begin{equation*}
\mathcal{C}(A)^{\bot_{\sigma}}=\big[\mathcal{C}_{1}^{\bot_{\sigma_{1}}},\mathcal{C}_{2}^{\bot_{\sigma_{2}}},\dots,\mathcal{C}_{k}^{\bot_{\sigma_{k}}}\big]\cdot B,
\end{equation*}
where $(B\otimes I_{n})M_{\tau}^{T}\big(\pi_{s}(A)^{T}\otimes I_{n}\big)
=\mathrm{diag}\big(M_{\widetilde{\tau}}^{T},M_{\widetilde{\tau}}^{T},\ldots,M_{\widetilde{\tau}}^{T}\big)$.

2) In 1), if $M_{\tau}=M_{\widehat{\tau}}\otimes M_{\widetilde{\tau}}$ for some $M_{\widehat{\tau}}$, then the $\sigma$ dual of $\mathcal{C}(A)$ is
\begin{equation*}
\mathcal{C}(A)^{\bot_{\sigma}}=\big[\mathcal{C}_{1}^{\bot_{\sigma_{1}}},\mathcal{C}_{2}^{\bot_{\sigma_{2}}},\dots,\mathcal{C}_{k}^{\bot_{\sigma_{k}}}\big]
\cdot\big(M_{\widehat{\tau}}^{T}\pi_{s}(A)^{T}\big)^{-1}.
\end{equation*}

3) In 2), if $M_{\widehat{\tau}}=I_{k}$, then $M_{\tau}=I_{k}\otimes M_{\widetilde{\tau}}$ and the $\sigma$ dual of $\mathcal{C}(A)$ is
\begin{equation*}
\mathcal{C}(A)^{\bot_{\sigma}}=\big[\mathcal{C}_{1}^{\bot_{\sigma_{1}}},\mathcal{C}_{2}^{\bot_{\sigma_{2}}},\dots,\mathcal{C}_{k}^{\bot_{\sigma_{k}}}\big]
\cdot\big(\pi_{s}(A)^{T}\big)^{-1}.
\end{equation*}

4) In 3), if $M_{\widetilde{\tau}}=I_{n}$ and $s=e-\ell$ with $0\leq \ell\leq e-1$, then one obtains the $\ell$-Galois dual of $\mathcal{C}(A)$, i.e.,
\begin{equation*}
\mathcal{C}(A)^{\bot_{\ell}}=\big[\mathcal{C}_{1}^{\bot_{\ell}},\mathcal{C}_{2}^{\bot_{\ell}},\dots,\mathcal{C}_{k}^{\bot_{\ell}}\big]
\cdot\big(\pi_{e-\ell}(A)^{T}\big)^{-1}.
\end{equation*}

5) In 4), if $\ell=0$, then one obtains the Euclidean hull of $\mathcal{C}(A)$, i.e.,
\begin{equation*}
\mathcal{C}(A)^{\bot_{E}}=\big[\mathcal{C}_{1}^{\bot_{E}},\mathcal{C}_{2}^{\bot_{E}},\ldots,\mathcal{C}_{k}^{\bot_{E}}\big]\cdot (A^{-1})^{T},
\end{equation*}
as shown in \cite[Proposition 6.2]{Blackmore2001} (see also Lemma \ref{lemma4.3}).

6) In 4), if $\ell=\frac{e}{2}$ for even $e$, then one obtains the Hermitian hull of $\mathcal{C}(A)$, i.e.,
\begin{equation*}
\mathcal{C}(A)^{\bot_{H}}=\big[\mathcal{C}_{1}^{\bot_{H}},\mathcal{C}_{2}^{\bot_{H}},\dots,\mathcal{C}_{k}^{\bot_{H}}\big]
\cdot\big(A^{-1}\big)^{\dag},
\end{equation*}
as shown in \cite[p.7]{Zhang2015Quantum}.
\end{remark}

\subsection{$\sigma$ hulls of matrix-product codes}

In the following theorem, we give a characterization on $\sigma$ hulls of matrix-product codes.

\begin{theorem}\label{theorem4.6}
Let $\mathcal{C}(A)=[\mathcal{C}_{1},\mathcal{C}_{2},\ldots,\mathcal{C}_{k}]\cdot A$,
where $\mathcal{C}_{i}$ is an $[n,m_{i}]_{q}$ code for $i=1,2,\ldots,k$ and $A\in\mathcal{M}(\mathbb{F}_{q},k\times r)$ with $k\leq r$.
Set $\sigma=(\tau,\pi_{s})\in\mathrm{SLAut}(\mathbb{F}_{q}^{rn})$ and $\sigma_{i}=(\tau_{i},\pi_{s})\in\mathrm{SLAut}(\mathbb{F}_{q}^{n})$,
where $\tau$ corresponds to a monomial matrix $M_{\tau}=D_{\tau}P_{\tau}\in\mathcal{M}(\mathbb{F}_{q},rn)$,
$\tau_{i}$ corresponds to a monomial matrix $M_{\tau_{i}}=D_{\tau_{i}}P_{\tau_{i}}\in\mathcal{M}(\mathbb{F}_{q},n)$ for $i=1,2,\ldots,k$, and $1\leq s\leq e$.
Suppose $A$ satisfies
\begin{equation*}
(A\otimes I_{n})M_{\tau}^{T}\big(\pi_{s}(A)^{T}\otimes I_{n}\big)
=\mathrm{diag}\big(\mu_{1}M_{\tau_{1}}^{T},\mu_{2}M_{\tau_{2}}^{T},\ldots,\mu_{k}M_{\tau_{k}}^{T}\big),
\end{equation*}
where $\mu_{1},\mu_{2},\ldots,\mu_{k}\in\mathbb{F}_{q}$. Then the $\sigma$ hull of $\mathcal{C}(A)$ is
\begin{equation}\label{eq4.2}
Hull_{\sigma}\big(\mathcal{C}(A)\big)=[\mathcal{B}_{1},\mathcal{B}_{2},\ldots,\mathcal{B}_{k}]\cdot A \ with \ \mathcal{B}_{i}
=\begin{cases}
\mathcal{C}_{i}, \ \ \ \ \ \ \ \ \ \ \ if \ \mu_{i}=0;\\
Hull_{\sigma_{i}}(\mathcal{C}_{i}), \ if \ \mu_{i}\neq0.
\end{cases}
\end{equation}
\end{theorem}

\begin{proof}
(1) First, let us prove $[\mathcal{B}_{1},\mathcal{B}_{2},\ldots,\mathcal{B}_{k}]\cdot A\subseteq Hull_{\sigma}\big(\mathcal{C}(A)\big)$.
For each $\mathbf{v}=[\mathbf{v}_{1},\mathbf{v}_{2},\ldots,\mathbf{v}_{k}]\cdot A\in[\mathcal{B}_{1},\mathcal{B}_{2},\ldots,\mathcal{B}_{k}]\cdot A$
with $\mathbf{v}_{i}\in\mathcal{B}_{i}$ for $i=1,2,\ldots,k$, we have $\mathbf{v}_{i}\in\mathcal{C}_{i}$ since $\mathcal{B}_{i}\subseteq\mathcal{C}_{i}$.
This leads to $\mathbf{v}\in\mathcal{C}(A)$.

Next, we are ready to prove $\mathbf{v}\in\mathcal{C}(A)^{\bot_{\sigma}}$.
For each $\mathbf{u}=[\mathbf{u}_{1},\mathbf{u}_{2},\ldots,\mathbf{u}_{k}]\cdot A\in\mathcal{C}(A)$ with $\mathbf{u}_{i}\in\mathcal{C}_{i}$ for $i=1,2,\ldots,k$,
it is calculated that
\begin{align*}
\langle\mathbf{v},\mathbf{u}\rangle_{\sigma}
&=\mathbf{v}\sigma(\mathbf{u})^{T}\\
&=\big([\mathbf{v}_{1},\mathbf{v}_{2},\ldots,\mathbf{v}_{k}](A\otimes I_{n})\big)
\sigma\big([\mathbf{u}_{1},\mathbf{u}_{2},\ldots,\mathbf{u}_{k}](A\otimes I_{n})\big)^{T}\\
&=\big([\mathbf{v}_{1},\mathbf{v}_{2},\ldots,\mathbf{v}_{k}](A\otimes I_{n})\big)
\Big(\pi_{s}\big([\mathbf{u}_{1},\mathbf{u}_{2},\ldots,\mathbf{u}_{k}](A\otimes I_{n})\big)M_{\tau}\Big)^{T}\\
&=\big([\mathbf{v}_{1},\mathbf{v}_{2},\ldots,\mathbf{v}_{k}](A\otimes I_{n})\big)
\Big([\pi_{s}(\mathbf{u}_{1}),\pi_{s}(\mathbf{u}_{2}),\ldots,\pi_{s}(\mathbf{u}_{k})]\big(\pi_{s}(A)\otimes I_{n}\big)M_{\tau}\Big)^{T}\\
&=[\mathbf{v}_{1},\mathbf{v}_{2},\ldots,\mathbf{v}_{k}]\Big((A\otimes I_{n})M_{\tau}^{T}\big(\pi_{s}(A)^{T}\otimes I_{n}\big)\Big)
[\pi_{s}(\mathbf{u}_{1}),\pi_{s}(\mathbf{u}_{2}),\ldots,\pi_{s}(\mathbf{u}_{k})]^{T}\\
&=[\mathbf{v}_{1},\mathbf{v}_{2},\ldots,\mathbf{v}_{k}]\mathrm{diag}\big(\mu_{1}M_{\tau_{1}}^{T},\mu_{2}M_{\tau_{2}}^{T},\ldots,\mu_{k}M_{\tau_{k}}^{T}\big)
[\pi_{s}(\mathbf{u}_{1}),\pi_{s}(\mathbf{u}_{2}),\ldots,\pi_{s}(\mathbf{u}_{k})]^{T}\\
&=\sum_{i=1}^{k}\mu_{i}\mathbf{v}_{i}\big(\pi_{s}(\mathbf{u}_{i})M_{\tau_{i}}\big)^{T}\\
&=\sum_{i=1}^{k}\mu_{i}\langle\mathbf{v}_{i},\mathbf{u}_{i}\rangle_{\sigma_{i}}\\
&=\sum_{1\leq i\leq k,\mu_{i}\neq 0}\mu_{i}\langle\mathbf{v}_{i},\mathbf{u}_{i}\rangle_{\sigma_{i}}\\
&=0,
\end{align*}
where the last equality follows from the fact $\mathbf{v}_{i}\in\mathcal{C}_{i}^{\bot_{\sigma_{i}}}$ if $\mu_{i}\neq 0$.
This reveals that $\mathbf{v}\in\mathcal{C}(A)^{\bot_{\sigma}}$.
So $\mathbf{v}\in Hull_{\sigma}\big(\mathcal{C}(A)\big)$ and thus
$[\mathcal{B}_{1},\mathcal{B}_{2},\ldots,\mathcal{B}_{k}]\cdot A\subseteq Hull_{\sigma}\big(\mathcal{C}(A)\big)$.

\vspace{6pt}

(2) We are now ready to prove $Hull_{\sigma}\big(\mathcal{C}(A)\big)\subseteq[\mathcal{B}_{1},\mathcal{B}_{2},\ldots,\mathcal{B}_{k}]\cdot A$.
For each $\mathbf{v}\in Hull_{\sigma}\big(\mathcal{C}(A)\big)$, it can be written as
$\mathbf{v}=[\mathbf{v}_{1},\mathbf{v}_{2},\ldots,\mathbf{v}_{k}]\cdot A$ with $\mathbf{v}_{i}\in\mathcal{C}_{i}$ for $i=1,2,\ldots,k$.
To achieve our goal, let us show that $\mathbf{v}_{i}\in\mathcal{B}_{i}$ for $i=1,2,\ldots,k$ as follows.

\vspace{4pt}

{\bfseries Case (i)}: Suppose $\mu_{j}=0$ for some $j$, then $\mathbf{v}_{j}\in\mathcal{C}_{j}=\mathcal{B}_{j}$.

\vspace{4pt}

{\bfseries Case (ii)}: Suppose $\mu_{j}\neq 0$ for some $j$. For each $\mathbf{u}_{j}\in\mathcal{C}_{j}$,
denote $\mathbf{u}=[\mathbf{0},\ldots,\mathbf{0},\mathbf{u}_{j},\mathbf{0},\ldots,\mathbf{0}]\cdot A\in\mathcal{C}(A)$.
By the analysis in part (1) and the fact $\mathbf{v}\in\mathcal{C}(A)^{\bot_{\sigma}}$, we have
$0=\langle\mathbf{v},\mathbf{u}\rangle_{\sigma}=\mu_{j}\langle\mathbf{v}_{j},\mathbf{u}_{j}\rangle_{\sigma_{j}}$,
i.e., $\langle\mathbf{v}_{j},\mathbf{u}_{j}\rangle_{\sigma_{j}}=0$. Then $\mathbf{v}_{j}\in\mathcal{C}_{j}^{\bot_{\sigma_{j}}}$ and thus
$\mathbf{v}_{j}\in Hull_{\sigma_{j}}(\mathcal{C}_{j})=\mathcal{B}_{j}$.

Consequently, $\mathbf{v}\in[\mathcal{B}_{1},\mathcal{B}_{2},\ldots,\mathcal{B}_{k}]\cdot A$.
So $Hull_{\sigma}\big(\mathcal{C}(A)\big)\subseteq[\mathcal{B}_{1},\mathcal{B}_{2},\ldots,\mathcal{B}_{k}]\cdot A$. $\hfill\square$
\end{proof}

\begin{remark}\label{remark4.7}
1) In Theorem \ref{theorem4.6}, if $\tau_{1}=\ldots=\tau_{k}=\widetilde{\tau}$ for some $\widetilde{\tau}$, then the $\sigma$ hull of $\mathcal{C}(A)$ is
given by Eq. (\ref{eq4.2}), where $(A\otimes I_{n})M_{\tau}^{T}\big(\pi_{s}(A)^{T}\otimes I_{n}\big)
=\mathrm{diag}\big(\mu_{1}M_{\widetilde{\tau}}^{T},\mu_{2}M_{\widetilde{\tau}}^{T},\ldots,\mu_{k}M_{\widetilde{\tau}}^{T}\big)$.

2) In 1), if $M_{\tau}=M_{\widehat{\tau}}\otimes M_{\widetilde{\tau}}$ for some $M_{\widehat{\tau}}$, then the $\sigma$ hull of $\mathcal{C}(A)$ is
given by Eq. (\ref{eq4.2}), where $AM_{\widehat{\tau}}^{T}\pi_{s}(A)^{T}=\mathrm{diag}(\mu_{1},\mu_{2},\ldots,\mu_{k})$.

3) In 2), if $M_{\widehat{\tau}}=I_{r}$, then $M_{\tau}=I_{r}\otimes M_{\widetilde{\tau}}$ and the $\sigma$ hull of $\mathcal{C}(A)$ is
given by Eq. (\ref{eq4.2}), where $A\pi_{s}(A)^{T}=\mathrm{diag}(\mu_{1},\mu_{2},\ldots,\mu_{k})$.

4) In 3), if $M_{\widetilde{\tau}}=I_{n}$ and $s=e-\ell$ with $0\leq \ell\leq e-1$, then one obtains the $\ell$-Galois hull of $\mathcal{C}(A)$, i.e.,
\begin{equation*}
Hull_{\ell}\big(\mathcal{C}(A)\big)=[\mathcal{B}_{1},\mathcal{B}_{2},\ldots,\mathcal{B}_{k}]\cdot A \ with \ \mathcal{B}_{i}
=\begin{cases}
\mathcal{C}_{i}, \ \ \ \ \ \ \ \ \ \ if \ \mu_{i}=0;\\
Hull_{\ell}(\mathcal{C}_{i}), \ if \ \mu_{i}\neq0,
\end{cases}
\end{equation*}
where $A\pi_{e-\ell}(A)^{T}=\mathrm{diag}(\mu_{1},\mu_{2},\ldots,\mu_{k})$, as shown in \cite[Theorem 5.6]{Liuhongwei2020Galois}.

5) In 4), if $\ell=0$, then one obtains the Euclidean hull of $\mathcal{C}(A)$, i.e.,
\begin{equation*}
Hull_{E}\big(\mathcal{C}(A)\big)=[\mathcal{B}_{1},\mathcal{B}_{2},\ldots,\mathcal{B}_{k}]\cdot A \ with \ \mathcal{B}_{i}
=\begin{cases}
\mathcal{C}_{i}, \ \ \ \ \ \ \ \ \ \ \ if \ \mu_{i}=0;\\
Hull_{E}(\mathcal{C}_{i}), \ if \ \mu_{i}\neq0,
\end{cases}
\end{equation*}
where $AA^{T}=\mathrm{diag}(\mu_{1},\mu_{2},\ldots,\mu_{k})$.

6) In 4), if $\ell=\frac{e}{2}$ for even $e$, then one obtains the Hermitian hull of $\mathcal{C}(A)$, i.e.,
\begin{equation*}
Hull_{H}\big(\mathcal{C}(A)\big)=[\mathcal{B}_{1},\mathcal{B}_{2},\ldots,\mathcal{B}_{k}]\cdot A \ with \ \mathcal{B}_{i}
=\begin{cases}
\mathcal{C}_{i}, \ \ \ \ \ \ \ \ \ \ \ if \ \mu_{i}=0;\\
Hull_{H}(\mathcal{C}_{i}), \ if \ \mu_{i}\neq0,
\end{cases}
\end{equation*}
where $AA^{\dag}=\mathrm{diag}(\mu_{1},\mu_{2},\ldots,\mu_{k})$.
\end{remark}

\begin{remark}\label{remark4.8}
On Remark \ref{remark4.7} 5), we make the following discussions:

1) In Remark \ref{remark4.7} 5), if $\mathcal{C}_{i}\subseteq\mathcal{C}_{i}^{\bot_{E}}$ for $i=1,2,\ldots,k$,
then $\mathcal{B}_{i}=\mathcal{C}_{i}$ and thus
{\setlength\abovedisplayskip{0.15cm}
\setlength\belowdisplayskip{0.15cm}
\begin{equation*}
Hull_{E}\big(\mathcal{C}(A)\big)=[\mathcal{C}_{1},\mathcal{C}_{2},\ldots,\mathcal{C}_{k}]\cdot A=\mathcal{C}(A),
\end{equation*}}revealing that $\mathcal{C}(A)\subseteq\mathcal{C}(A)^{\bot_{E}}$.
This gives a new manner for deriving the result in \cite[Theorem III.1]{Mankean2016Matrixproduct}.

2) In Remark \ref{remark4.7} 5), if $k=r$, $\mathcal{C}_{i}^{\bot_{E}}\subseteq\mathcal{C}_{i}$ and $\mu_{i}\neq 0$ for $i=1,2,\ldots,k$,
then $\mathcal{B}_{i}=\mathcal{C}_{i}^{\bot_{E}}$ and thus
{\setlength\abovedisplayskip{0.15cm}
\setlength\belowdisplayskip{0.15cm}
\begin{align*}
Hull_{E}\big(\mathcal{C}(A)\big)
&=\big[\mathcal{C}_{1}^{\bot_{E}},\mathcal{C}_{2}^{\bot_{E}},\ldots,\mathcal{C}_{k}^{\bot_{E}}\big]\cdot \mathrm{diag}(\mu_{1},\mu_{2},\ldots,\mu_{k})(A^{-1})^{T}\\
&=\mathcal{C}(A)^{\bot_{E}},
\end{align*}}which reveals that $\mathcal{C}(A)^{\bot_{E}}\subseteq\mathcal{C}(A)$.

3) In Remark \ref{remark4.7} 5), if $k=r$ and $A\in\mathcal{M}(\mathbb{F}_{q},k)$ is invertible, then by \cite[Theorem 2]{Cao2020New},
there exists an invertible matrix $N\in\mathcal{M}(\mathbb{F}_{q},k)$ such that $N^{T}A$ is a quasi-orthogonal matrix,
i.e., $N^{T}A(N^{T}A)^{T}$ is an invertible diagonal matrix. From Remark \ref{remark4.7} 5), the Euclidean hull of $\mathcal{C}(N^{T}A)$ is
{\setlength\abovedisplayskip{0.15cm}
\setlength\belowdisplayskip{0.15cm}
\begin{equation*}
Hull_{E}\big(\mathcal{C}(N^{T}A)\big)=\big[Hull_{E}(\mathcal{C}_{1}),Hull_{E}(\mathcal{C}_{2}),\ldots,Hull_{E}(\mathcal{C}_{k})\big]\cdot N^{T}A.
\end{equation*}}
\end{remark}

\begin{remark}\label{remark4.9}
On Remark \ref{remark4.7} 6), we make the following discussions:

1) In Remark \ref{remark4.7} 6), if $\mathcal{C}_{i}\subseteq\mathcal{C}_{i}^{\bot_{H}}$ for $i=1,2,\ldots,k$,
then $\mathcal{B}_{i}=\mathcal{C}_{i}$ and thus
{\setlength\abovedisplayskip{0.15cm}
\setlength\belowdisplayskip{0.15cm}
\begin{equation*}
Hull_{H}\big(\mathcal{C}(A)\big)=[\mathcal{C}_{1},\mathcal{C}_{2},\ldots,\mathcal{C}_{k}]\cdot A=\mathcal{C}(A),
\end{equation*}}revealing that $\mathcal{C}(A)\subseteq\mathcal{C}(A)^{\bot_{H}}$.
This gives a new manner for deriving the result in \cite[Theorem 3.1]{Jitman2017}.

2) In Remark \ref{remark4.7} 6), if $k=r$, $\mathcal{C}_{i}^{\bot_{H}}\subseteq\mathcal{C}_{i}$ and $\mu_{i}\neq 0$ for $i=1,2,\ldots,k$,
then $\mathcal{B}_{i}=\mathcal{C}_{i}^{\bot_{H}}$ and thus
{\setlength\abovedisplayskip{0.15cm}
\setlength\belowdisplayskip{0.15cm}
\begin{align*}
Hull_{H}\big(\mathcal{C}(A)\big)
&=\big[\mathcal{C}_{1}^{\bot_{H}},\mathcal{C}_{2}^{\bot_{H}},\ldots,\mathcal{C}_{k}^{\bot_{H}}\big]\cdot \mathrm{diag}(\mu_{1},\mu_{2},\ldots,\mu_{k})(A^{-1})^{\dag}\\
&=\mathcal{C}(A)^{\bot_{H}},
\end{align*}}revealing that $\mathcal{C}(A)^{\bot_{H}}\subseteq\mathcal{C}(A)$.
This gives a new manner for deriving the result in \cite[Theorem 3.1]{Liu2018On}.

3) In Remark \ref{remark4.7} 6), if $k=r$ and $A\in\mathcal{M}(\mathbb{F}_{q},k)$ is invertible,
then by \cite[Theorem 1.7.10]{Giuzzi2000Hermitian} there exists an invertible matrix $U\in\mathcal{M}(\mathbb{F}_{q},k)$ such that $UA$ is a quasi-unitary matrix,
i.e., $UA(UA)^{\dag}$ is an invertible diagonal matrix. From Remark \ref{remark4.7} 6), the Hermitian hull of $\mathcal{C}(UA)$ is
{\setlength\abovedisplayskip{0.15cm}
\setlength\belowdisplayskip{0.15cm}
\begin{equation*}
Hull_{H}\big(\mathcal{C}(UA)\big)=\big[Hull_{H}(\mathcal{C}_{1}),Hull_{H}(\mathcal{C}_{2}),\ldots,Hull_{H}(\mathcal{C}_{k})\big]\cdot UA.
\end{equation*}}
\end{remark}

\subsection{Intersection of matrix-product codes}

Let us give a characterization on the intersection of a pair of matrix-product codes.

\begin{theorem}\label{theorem4.10}
Let $A,B\in\mathcal{M}(\mathbb{F}_{q},k)$. Define a pair of matrix-product codes as follows:
\begin{equation*}
\mathcal{C}(A)=[\mathcal{C}_{1},\mathcal{C}_{2},\ldots,\mathcal{C}_{k}]\cdot A,\ \ \ \ \mathcal{D}(B)=[\mathcal{D}_{1},\mathcal{D}_{2},\ldots,\mathcal{D}_{k}]\cdot B,
\end{equation*}
where $\mathcal{C}_{i}$ and $\mathcal{D}_{i}$ are linear codes of length $n$ for $i=1,2,\ldots,k$.
Suppose $A$ is invertible and $BA^{-1}=\mathrm{diag}(\mu_{1},\mu_{2},\ldots,\mu_{k})$ with $\mu_{1},\mu_{2},\ldots,\mu_{k}\in\mathbb{F}_{q}$.
Then the intersection of $\mathcal{C}(A)$ and $\mathcal{D}(B)$ is
\begin{equation*}
\mathcal{C}(A)\cap\mathcal{D}(B)=[\mathcal{B}_{1},\mathcal{B}_{2},\ldots,\mathcal{B}_{k}]\cdot B \ with \ \mathcal{B}_{i}
=\begin{cases}
\mathcal{D}_{i}, \ \ \ \ \ \ \ \ if \ \mu_{i}=0;\\
\mathcal{C}_{i}\cap\mathcal{D}_{i}, \ \ \ if \ \mu_{i}\neq0.
\end{cases}
\end{equation*}
\end{theorem}

\begin{proof}
(1) First, let us prove $[\mathcal{B}_{1},\mathcal{B}_{2},\ldots,\mathcal{B}_{k}]\cdot B\subseteq\mathcal{C}(A)\cap\mathcal{D}(B)$.
For each $\mathbf{v}=[\mathbf{v}_{1},\mathbf{v}_{2},\ldots,\mathbf{v}_{k}]\cdot B\in[\mathcal{B}_{1},\mathcal{B}_{2},\ldots,\mathcal{B}_{k}]\cdot B$ with
$\mathbf{v}_{i}\in\mathcal{B}_{i}$ for $i=1,2,\ldots,k$, we have $\mathbf{v}_{i}\in\mathcal{D}_{i}$ since $\mathcal{B}_{i}\subseteq\mathcal{D}_{i}$
for $i=1,2,\ldots,k$. Hence $\mathbf{v}\in\mathcal{D}(B)$.
We claim that $\mu_{i}\mathbf{v}_{i}\in\mathcal{C}_{i}$ for $i=1,2,\ldots,k$, because

\vspace{4pt}

\ \ {\bfseries Case (i)}: If $\mu_{j}=0$ for some $j$, then $\mu_{j}\mathbf{v}_{j}=\mathbf{0}\in\mathcal{C}_{j}$.

\vspace{4pt}

\ \ {\bfseries Case (ii)}: If $\mu_{j}\neq 0$ for some $j$, then $\mu_{j}\mathbf{v}_{j}\in\mathcal{B}_{j}
=\mathcal{C}_{j}\cap\mathcal{D}_{j}\subseteq\mathcal{C}_{j}$.

So $\mu_{i}\mathbf{v}_{i}\in\mathcal{C}_{i}$ for $i=1,2,\ldots,k$.  Then it follows that
\begin{align*}
\mathbf{v}
&=[\mathbf{v}_{1},\mathbf{v}_{2},\ldots,\mathbf{v}_{k}]\cdot B\\
&=[\mathbf{v}_{1},\mathbf{v}_{2},\ldots,\mathbf{v}_{k}]\cdot \mathrm{diag}(\mu_{1},\mu_{2},\ldots,\mu_{k})A\\
&=[\mathbf{v}_{1},\mathbf{v}_{2},\ldots,\mathbf{v}_{k}]\big(\mathrm{diag}(\mu_{1},\mu_{2},\ldots,\mu_{k})\otimes I_{n}\big)(A\otimes I_{n})\\
&=[\mathbf{v}_{1},\mathbf{v}_{2},\ldots,\mathbf{v}_{k}]\mathrm{diag}(\mu_{1}I_{n},\mu_{2}I_{n},\ldots,\mu_{k}I_{n})(A\otimes I_{n})\\
&=[\mu_{1}\mathbf{v}_{1},\mu_{2}\mathbf{v}_{2},\ldots,\mu_{k}\mathbf{v}_{k}](A\otimes I_{n})\\
&\in[\mathcal{C}_{1},\mathcal{C}_{2},\ldots,\mathcal{C}_{k}](A\otimes I_{n})=\mathcal{C}(A).
\end{align*}
Consequently, $\mathbf{v}\in\mathcal{C}(A)\cap\mathcal{D}(B)$ and thus
$[\mathcal{B}_{1},\mathcal{B}_{2},\ldots,\mathcal{B}_{k}]\cdot B\subseteq\mathcal{C}(A)\cap\mathcal{D}(B)$.

\vspace{6pt}

(2) We are now ready to prove $\mathcal{C}(A)\cap\mathcal{D}(B)\subseteq[\mathcal{B}_{1},\mathcal{B}_{2},\ldots,\mathcal{B}_{k}]\cdot B$.
For each $\mathbf{v}\in\mathcal{C}(A)\cap\mathcal{D}(B)$, we have $\mathbf{v}\in\mathcal{D}(B)$ and thus it can be written as
$\mathbf{v}=[\mathbf{v}_{1},\mathbf{v}_{2},\ldots,\mathbf{v}_{k}]\cdot B$ with $\mathbf{v}_{i}\in\mathcal{D}_{i}$ for $i=1,2,\ldots,k$.

We claim that $\mathbf{v}_{i}\in\mathcal{B}_{i}$ for $i=1,2,\ldots,k$. This is because

\vspace{4pt}

\ \ {\bfseries Case (a)}: Suppose $\mu_{j}=0$ for some $j$, then $\mathbf{v}_{j}\in\mathcal{D}_{j}=\mathcal{B}_{j}$.

\vspace{4pt}

\ \ {\bfseries Case (b)}: Suppose $\mu_{j}\neq 0$ for some $j$. Note that
$\mathbf{v}=[\mu_{1}\mathbf{v}_{1},\mu_{2}\mathbf{v}_{2},\ldots,\mu_{k}\mathbf{v}_{k}]\cdot A$, as shown in part (1).
Besides, we have $\mathbf{v}\in\mathcal{C}(A)=[\mathcal{C}_{1},\mathcal{C}_{2},\ldots,\mathcal{C}_{k}]\cdot A$. Thus
$[\mu_{1}\mathbf{v}_{1},\mu_{2}\mathbf{v}_{2},\ldots,\mu_{k}\mathbf{v}_{k}]\cdot A\in [\mathcal{C}_{1},\mathcal{C}_{2},\ldots,\mathcal{C}_{k}]\cdot A$.
Since $A$ is invertible, we obtain $\mu_{j}\mathbf{v}_{j}\in\mathcal{C}_{j}$, i.e., $\mathbf{v}_{j}\in\mathcal{C}_{j}$,
which implies that $\mathbf{v}_{j}\in\mathcal{C}_{j}\cap\mathcal{D}_{j}=\mathcal{B}_{j}$.

By the above cases, we obtain
$\mathbf{v}_{i}\in\mathcal{B}_{i}$ for $i=1,2,\ldots,k$, and thus $\mathbf{v}\in[\mathcal{B}_{1},\mathcal{B}_{2},\ldots,\mathcal{B}_{k}]\cdot B$.
Therefore, $\mathcal{C}(A)\cap\mathcal{D}(B)\subseteq[\mathcal{B}_{1},\mathcal{B}_{2},\ldots,\mathcal{B}_{k}]\cdot B$.

By part (1) and part (2), we complete the proof. $\hfill\square$
\end{proof}

\section{MDS codes with new $\ell$-Galois hulls}\label{section5}
In this section, we will give a necessary and sufficient condition under which any codeword of a GRS code
or an extended GRS code is contained in its $\sigma$ dual,
generalizing those for Euclidean case in \cite{Chen2018New}, Hermitian case in \cite{Fang2018Two} and $\ell$-Galois case in \cite{Cao2021MDS}.
As an application, we will construct eleven families of $q$-ary MDS codes with new $\ell$-Galois hulls that are not covered
by the latest papers \cite{Cao2021MDS} and \cite{Fang2022New}.

\subsection{Codewords of $\sigma$ hulls of GRS codes and extended GRS codes}\label{subsection5.1}

Let $a_{1},a_{2},\ldots,a_{n}$ be $n$ distinct elements in $\mathbb{F}_{q}$ and let $v_{1},v_{2},\ldots,v_{n}\in \mathbb{F}_{q}^{\ast}$.
Put $\mathbf{a}=(a_{1},a_{2},\ldots,a_{n})$ and $\mathbf{v}=(v_{1},v_{2},\ldots,v_{n})$.
The \emph{generalized Reed-Solomon (GRS) code} associated with $\mathbf{a}$ and $\mathbf{v}$ is defined as
{\setlength\abovedisplayskip{0.15cm}
\setlength\belowdisplayskip{0.15cm}
\begin{align*}
GRS_{k}(\mathbf{a},\mathbf{v})=\big\{\big(v_{1}f(a_{1}),v_{2}f(a_{2}),\ldots,v_{n}f(a_{n})\big)\big|f(x)\in \mathbb{F}_{q}[x], \mathrm{deg}\big(f(x)\big)\leq k-1\big\}.
\end{align*}}It is an $[n,k]_{q}$ MDS code with a generator matrix:
\begin{align}\label{eq5.1}
G_{k}(\mathbf{a},\mathbf{v})=\left[
\begin{array}{cccc}
v_{1}& v_{2}&  \cdots &v_{n}\\[4pt]
v_{1}a_{1}&v_{2}a_{2}& \cdots &v_{n}a_{n}\\
\vdots&\vdots&\ddots&\vdots\\
v_{1}a_{1}^{k-1}& v_{2}a_{2}^{k-1}& \cdots &v_{n}a_{n}^{k-1}  \\
\end{array}\right].
\end{align}
Moreover, the \emph{extended GRS code} associated with $\mathbf{a}$ and $\mathbf{v}$ is defined as
\begin{align*}
GRS_{k}(\mathbf{a},\mathbf{v},\infty)=\big\{\big(v_{1}f(a_{1}),\ldots,v_{n}f(a_{n}),f_{k-1}\big)\big|f(x)\in \mathbb{F}_{q}[x],
\mathrm{deg}\big(f(x)\big)\leq k-1\big\},
\end{align*}
where $f_{k-1}$ denotes the coefficient of $x^{k-1}$ in $f(x)$. It is not difficult to verify that $GRS_{k}(\mathbf{a},\mathbf{v},\infty)$
is an $[n+1,k]_{q}$ MDS code with a generator matrix:
\begin{align}\label{eq5.2}
G_{k}(\mathbf{a},\mathbf{v},\infty)=\left[
\begin{array}{ccccc}
v_{1}& v_{2}&  \cdots &v_{n}&0\\[4pt]
v_{1}a_{1}&v_{2}a_{2}& \cdots &v_{n}a_{n}&0\\
\vdots&\vdots&\ddots&\vdots&\vdots\\
v_{1}a_{1}^{k-2}& v_{2}a_{2}^{k-2}& \cdots &v_{n}a_{n}^{k-2}&0  \\[4pt]
v_{1}a_{1}^{k-1}& v_{2}a_{2}^{k-1}& \cdots &v_{n}a_{n}^{k-1}&1  \\
\end{array}\right].
\end{align}

From now on, for $i=1,2,\ldots,n$, we shall denote
\begin{align}\label{eq5.3}
u_{i}:=\prod_{1\leq j\leq n,j\neq i}(a_{i}-a_{j})^{-1}.
\end{align}

Denote by $\mathbf{1}=(1,1,\ldots,1)$ the all one vector. The Euclidean dual $GRS_{k}(\mathbf{a},\mathbf{1})^{\perp_{E}}$
of $GRS_{k}(\mathbf{a},\mathbf{1})$ and the Euclidean dual $GRS_{k}(\mathbf{a},\mathbf{1},\infty)^{\perp_{E}}$ of $GRS_{k}(\mathbf{a},\mathbf{1},\infty)$
have the following expressions.

\begin{lemma}\label{lemma5.1}
{\rm{(\!\!\cite[Lemma 2.3 (i)]{Jin2016New})}}
Let $\mathbf{a}=(a_{1},a_{2},\ldots,a_{n})$ and $\mathbf{u}=(u_{1},u_{2},\ldots,u_{n})$,
where $a_{1},a_{2},\ldots,a_{n}$ are $n$ distinct elements in $\mathbb{F}_{q}$ and $u_{i}$ is defined by Eq. (\ref{eq5.3}) for $i=1,2,\ldots,n$.
Then
{\setlength\abovedisplayskip{0.15cm}
\setlength\belowdisplayskip{0.15cm}
\begin{equation*}
GRS_{k}(\mathbf{a},\mathbf{1})^{\perp_{E}}=GRS_{n-k}(\mathbf{a},\mathbf{u}).
\end{equation*}}
\end{lemma}

\begin{lemma}\label{lemma5.2}
{\rm{(\!\!\cite[Lemma 5]{Fang2018Two})}}
Let $\mathbf{a}=(a_{1},a_{2},\ldots,a_{n})$ and $\mathbf{u}=(u_{1},u_{2},\ldots,u_{n})$,
where $a_{1},a_{2},\ldots,a_{n}$ are $n$ distinct elements in $\mathbb{F}_{q}$ and $u_{i}$ is defined by Eq. (\ref{eq5.3}) for $i=1,2,\ldots,n$.
Then
{\setlength\abovedisplayskip{0.15cm}
\setlength\belowdisplayskip{0.15cm}
\begin{align*}
GRS_{k}(\mathbf{a},\mathbf{1},\infty)^{\perp_{E}}=\big\{\big(u_{1}g(a_{1}),\ldots,u_{n}g(a_{n}),-g_{n-k}\big)
\big|g(x)\in \mathbb{F}_{q}[x],\mathrm{deg}\big(g(x)\big)\leq n-k\big\},
\end{align*}}where $g_{n-k}$ denotes the coefficient of $x^{n-k}$ in $g(x)$.
\end{lemma}

The results in Lemmas \ref{lemma5.1} and \ref{lemma5.2} can be extended to the following forms.

\begin{lemma}\label{lemma5.3}
Let $s$ be an integer with $1\leq s\leq e$. Then
\begin{equation*}
\Big(\pi_{s}\big(GRS_{k}(\mathbf{a},\mathbf{1})\big)\Big)^{\perp_{E}}=\pi_{s}\big(GRS_{n-k}(\mathbf{a},\mathbf{u})\big).
\end{equation*}
\end{lemma}

\begin{proof}
Clearly, $\mathrm{dim}_{\mathbb{F}_{q}}\pi_{s}\big(GRS_{n-k}(\mathbf{a},\mathbf{u})\big)
=\mathrm{dim}_{\mathbb{F}_{q}}\Big(\pi_{s}\big(GRS_{k}(\mathbf{a},\mathbf{1})\big)\Big)^{\perp_{E}}$.
For any $\mathbf{c}\in GRS_{k}(\mathbf{a},\mathbf{1})$ and
$\mathbf{d}\in GRS_{n-k}(\mathbf{a},\mathbf{u})$,
we have $\langle\pi_{s}(\mathbf{c}),\pi_{s}(\mathbf{d})\rangle_{E}=\pi_{s}\big(\langle\mathbf{c},\mathbf{d}\rangle_{E}\big)=0$ by Lemma \ref{lemma5.1}.
Thus,
$\pi_{s}\big(GRS_{n-k}(\mathbf{a},\mathbf{u})\big)\subseteq\Big(\pi_{s}\big(GRS_{k}(\mathbf{a},\mathbf{1})\big)\Big)^{\perp_{E}}$.
Hence $\pi_{s}\big(GRS_{n-k}(\mathbf{a},\mathbf{u})\big)=\Big(\pi_{s}\big(GRS_{k}(\mathbf{a},\mathbf{1})\big)\Big)^{\perp_{E}}$. $\hfill\square$
\end{proof}

\begin{lemma}\label{lemma5.4}
Let $s$ be an integer with $1\leq s\leq e$. Then
\begin{equation*}
\Big(\pi_{s}\big(GRS_{k}(\mathbf{a},\mathbf{1},\infty)\big)\Big)^{\perp_{E}}
=\Big\{\Big(\pi_{s}\big(u_{1}g(a_{1})\big),\ldots,\pi_{s}\big(u_{n}g(a_{n})\big),-\pi_{s}(g_{n-k})\Big)
\Big|g(x)\in \mathbb{F}_{q}[x],\mathrm{deg}\big(g(x)\big)\leq n-k\Big\},
\end{equation*}
where $g_{n-k}$ denotes the coefficient of $x^{n-k}$ in $g(x)$.
\end{lemma}

\begin{proof}
First, it is easy to check that both sides are $(n-k+1)$-dimensional linear spaces over $\mathbb{F}_{q}$.
Besides, for any $\mathbf{c}=\big(f(a_{1}),\ldots,f(a_{n}),f_{k-1}\big)\in GRS_{k}(\mathbf{a},\mathbf{1},\infty)$ with $\mathrm{deg}\big(f(x)\big)\leq k-1$ and
$\mathbf{d}=\Big(\pi_{s}\big(u_{1}g(a_{1})\big),\ldots,\pi_{s}\big(u_{n}g(a_{n})\big),-\pi_{s}(g_{n-k})\Big)$
with $\mathrm{deg}\big(g(x)\big)\leq n-k$, it follows from Lemma \ref{lemma5.2} that
{\setlength\abovedisplayskip{0.15cm}
\setlength\belowdisplayskip{0.15cm}
\begin{align*}
\langle\pi_{s}(\mathbf{c}),\mathbf{d}\rangle_{E}
&=\sum_{i=1}^{n}\pi_{s}\big(f(a_{i})\big)\pi_{s}\big(u_{i}g(a_{i})\big)-\pi_{s}(f_{k-1})\pi_{s}(g_{n-k})\\
&=\pi_{s}\Big(\sum_{i=1}^{n}f(a_{i})\cdot u_{i}g(a_{i})-f_{k-1}g_{n-k}\Big)=0,
\end{align*}}which reveals that the linear space on the right side is contained in $\Big(\pi_{s}\big(GRS_{k}(\mathbf{a},\mathbf{1},\infty)\big)\Big)^{\perp_{E}}$. $\hfill\square$
\end{proof}

\vspace{6pt}

Based on Lemma \ref{lemma5.3}, we give a necessary and sufficient condition under which any codeword of a GRS code is contained in its $\sigma$ dual
as follows.

\begin{proposition}\label{proposition5.5}
Set $\sigma=(\tau,\pi_{s})\in\mathrm{SLAut}(\mathbb{F}_{q}^{n})$, where $\tau$ corresponds to a monomial matrix
$M_{\tau}=D_{\tau}P_{\tau}\in\mathcal{M}(\mathbb{F}_{q},n)$
in which $D_{\tau}=\mathrm{diag}(\mu_{1},\mu_{2},\ldots,\mu_{n})$ with $\mu_{i}\in\mathbb{F}_{q}^{\ast}$ for $i=1,2,\ldots,n$, and $1\leq s\leq e$.
Then for any $\mathbf{c}=\big(v_{1}f(a_{1}),v_{2}f(a_{2}),\ldots,v_{n}f(a_{n})\big)\in GRS_{k}(\mathbf{a},\mathbf{v})$,
we have $\mathbf{c}\in GRS_{k}(\mathbf{a},\mathbf{v})^{\bot_{\sigma}}$ if and only if there exists a polynomial $g(x)\in \mathbb{F}_{q}[x]$ with $\mathrm{deg}\big(g(x)\big)\leq n-k-1$ such that
{\setlength\abovedisplayskip{0.15cm}
\setlength\belowdisplayskip{0.15cm}
\begin{align*}
\Big(\mu_{1}\pi_{s}(v_{1})v_{\tau^{-1}(1)}f\big(a_{\tau^{-1}(1)}\big),\ldots,\mu_{n}\pi_{s}(v_{n})v_{\tau^{-1}(n)}f\big(a_{\tau^{-1}(n)}\big)\Big)
=\Big(\pi_{s}\big(u_{1}g(a_{1})\big),\ldots,\pi_{s}\big(u_{n}g(a_{n})\big)\Big).
\end{align*}}
\end{proposition}

\begin{proof}
It follows from  Eq. (\ref{eq5.1}) that $G_{k}(\mathbf{a},\mathbf{v})=G_{k}(\mathbf{a},\mathbf{1})V$ with $V=\mathrm{diag}(v_{1},v_{2},\ldots,v_{n})$.
This together with Lemma \ref{lemma5.3} derives that
for any $\mathbf{c}=\big(v_{1}f(a_{1}),v_{2}f(a_{2}),\ldots,v_{n}f(a_{n})\big)\in GRS_{k}(\mathbf{a},\mathbf{v})$, one has
{\setlength\abovedisplayskip{0.15cm}
\setlength\belowdisplayskip{0.15cm}
\begin{align*}
\mathbf{c}\in GRS_{k}(\mathbf{a},\mathbf{v})^{\bot_{\sigma}}
&\Longleftrightarrow \mathbf{c}\in \Big(\pi_{s}\big(GRS_{k}(\mathbf{a},\mathbf{v})\big)D_{\tau}P_{\tau}\Big)^{\bot_{E}}\\
&\Longleftrightarrow \pi_{s}\big(G_{k}(\mathbf{a},\mathbf{v})\big)D_{\tau}P_{\tau}\mathbf{c}^{T}=\mathbf{0}\\
&\Longleftrightarrow \pi_{s}\big(G_{k}(\mathbf{a},\mathbf{1})\big)\pi_{s}(V)D_{\tau}P_{\tau}\mathbf{c}^{T}=\mathbf{0}\\
&\Longleftrightarrow \mathbf{c}P_{\tau}^{T}D_{\tau}\pi_{s}(V)\in \Big(\pi_{s}\big(GRS_{k}(\mathbf{a},\mathbf{1})\big)\Big)^{\bot_{E}}\\
&\Longleftrightarrow \mathbf{c}P_{\tau}^{T}D_{\tau}\pi_{s}(V)\in\pi_{s}\big(GRS_{n-k}(\mathbf{a},\mathbf{u})\big).
\end{align*}}That is, $\mathbf{c}\in GRS_{k}(\mathbf{a},\mathbf{v})^{\bot_{\sigma}}$ if and only if
{\setlength\abovedisplayskip{0.15cm}
\setlength\belowdisplayskip{0.15cm}
\begin{equation*}
\big(v_{1}f(a_{1}),\ldots,v_{n}f(a_{n})\big)P_{\tau}^{T}\mathrm{diag}(\mu_{1},\ldots,\mu_{n})
\mathrm{diag}\big(\pi_{s}(v_{1}),\ldots,\pi_{s}(v_{n})\big)\in\pi_{s}\big(GRS_{n-k}(\mathbf{a},\mathbf{u})\big).
\end{equation*}}By Eq. (\ref{eq2.2}), the desired result then follows. $\hfill\square$
\end{proof}

\begin{remark}\label{remark5.6}
Consider $P_{\tau}=I_{n}$ in Proposition \ref{proposition5.5}, then $\mathbf{c}\in Hull_{\sigma}\big(GRS_{k}(\mathbf{a},\mathbf{v})\big)$
if and only if there exists a polynomial $g(x)\in \mathbb{F}_{q}[x]$ with $\mathrm{deg}\big(g(x)\big)\leq n-k-1$ such that
\begin{align}\label{eq5.4}
\big(\mu_{1}\pi_{s}(v_{1})v_{1}f(a_{1}),\ldots,\mu_{n}\pi_{s}(v_{n})v_{n}f(a_{n})\big)
=\Big(\pi_{s}\big(u_{1}g(a_{1})\big),\ldots,\pi_{s}\big(u_{n}g(a_{n})\big)\Big).
\end{align}
Further, if $D_{\tau}=I_{n}$, then Eq. (\ref{eq5.4}) becomes
{\setlength\abovedisplayskip{0.15cm}
\setlength\belowdisplayskip{0.15cm}
\begin{align*}
\big(\pi_{s}(v_{1})v_{1}f(a_{1}),\ldots,\pi_{s}(v_{n})v_{n}f(a_{n})\big)
=\Big(\pi_{s}\big(u_{1}g(a_{1})\big),\ldots,\pi_{s}\big(u_{n}g(a_{n})\big)\Big),
\end{align*}}which contains the following cases:

$\bullet$ For $s=e$, one gets the Euclidean case shown in \cite[Lemma 2]{Chen2018New}:
{\setlength\abovedisplayskip{0.15cm}
\setlength\belowdisplayskip{0.15cm}
\begin{align*}
\big(v_{1}^{2}f(a_{1}),\ldots,v_{n}^{2}f(a_{n})\big)
=\big(u_{1}g(a_{1}),\ldots,u_{n}g(a_{n})\big).
\end{align*}

$\bullet$ For $s=\frac{e}{2}$ if $e$ is even, one gets the Hermitian case shown in \cite[Lemma 6]{Fang2018Two}:
\begin{align*}
\big(v_{1}^{\sqrt{q}+1}f^{\sqrt{q}}(a_{1}),\ldots,v_{n}^{\sqrt{q}+1}f^{\sqrt{q}}(a_{n})\big)
=\big(u_{1}g(a_{1}),\ldots,u_{n}g(a_{n})\big).
\end{align*}

$\bullet$ For $s=e-\ell$ with $0\leq \ell\leq e-1$, one gets the $\ell$-Galois case shown in \cite[Proposition II.1]{Cao2021MDS}:
\begin{align}\label{eq5.5}
\big(v_{1}^{p^{\ell}+1}f^{p^{\ell}}(a_{1}),\ldots,v_{n}^{p^{\ell}+1}f^{p^{\ell}}(a_{n})\big)
=\big(u_{1}g(a_{1}),\ldots,u_{n}g(a_{n})\big).
\end{align}}
\end{remark}

Next, by using Lemma \ref{lemma5.4}, we will give a necessary and sufficient condition under which any codeword of an extended GRS code
is contained in its $\sigma$ dual.

\begin{proposition}\label{proposition5.7}
Set $\sigma=(\tau,\pi_{s})\in\mathrm{SLAut}(\mathbb{F}_{q}^{n+1})$, where $\tau$ corresponds to a monomial matrix
$M_{\tau}=D_{\tau}P_{\tau}\in\mathcal{M}\big(\mathbb{F}_{q},n+1\big)$ in which
$D_{\tau}=\mathrm{diag}(\mu_{1},\mu_{2},\ldots,\mu_{n+1})$ with $\mu_{i}\in\mathbb{F}_{q}^{\ast}$ for $i=1,2,\ldots,n+1$, and $1\leq s\leq e$.
Then for any $\mathbf{c}=\big(v_{1}f(a_{1}),\ldots,v_{n}f(a_{n}),f_{k-1}\big)\in GRS_{k}(\mathbf{a},\mathbf{v},\infty)$,
we have $\mathbf{c}\in GRS_{k}(\mathbf{a},\mathbf{v},\infty)^{\bot_{\sigma}}$ if and only if there exists a polynomial $g(x)\in \mathbb{F}_{q}[x]$ with $\mathrm{deg}\big(g(x)\big)\leq n-k$ such that
\begin{align*}
\big(\mu_{1}\pi_{s}(v_{1})w_{\tau^{-1}(1)},\ldots,\mu_{n}\pi_{s}(v_{n})w_{\tau^{-1}(n)},\mu_{n+1}w_{\tau^{-1}(n+1)}\big)
=\Big(\pi_{s}\big(u_{1}g(a_{1})\big),\ldots,\pi_{s}\big(u_{n}g(a_{n})\big),-\pi_{s}(g_{n-k})\Big),
\end{align*}
\end{proposition}
where $w_{i}:=v_{i}f(a_{i})$ for $i=1,2,\ldots,n$ and $w_{n+1}:=f_{k-1}$.

\vspace{6pt}

\begin{proof}
In light of Eq. (\ref{eq5.2}), we obtain $G_{k}(\mathbf{a},\mathbf{v},\infty)=G_{k}(\mathbf{a},\mathbf{1},\infty)\widetilde{V}$ with $\widetilde{V}=\mathrm{diag}(v_{1},\ldots,v_{n},1)$.
Then for any $\mathbf{c}=\big(v_{1}f(a_{1}),\ldots,v_{n}f(a_{n}),f_{k-1}\big)\in GRS_{k}(\mathbf{a},\mathbf{v},\infty)$, one has
{\setlength\abovedisplayskip{0.15cm}
\setlength\belowdisplayskip{0.15cm}
\begin{align*}
\mathbf{c}\in GRS_{k}(\mathbf{a},\mathbf{v},\infty)^{\bot_{\sigma}}
&\Longleftrightarrow \mathbf{c}\in \Big(\pi_{s}\big(GRS_{k}(\mathbf{a},\mathbf{v},\infty)\big)D_{\tau}P_{\tau}\Big)^{\bot_{E}}\\
&\Longleftrightarrow \pi_{s}\big(G_{k}(\mathbf{a},\mathbf{v},\infty)\big)D_{\tau}P_{\tau}\mathbf{c}^{T}=\mathbf{0}\\
&\Longleftrightarrow \pi_{s}\big(G_{k}(\mathbf{a},\mathbf{1},\infty)\big)\pi_{s}(\widetilde{V})D_{\tau}P_{\tau}\mathbf{c}^{T}=\mathbf{0}\\
&\Longleftrightarrow \mathbf{c}P_{\tau}^{T}D_{\tau}\pi_{s}(\widetilde{V})\in \Big(\pi_{s}\big(GRS_{k}(\mathbf{a},\mathbf{1},\infty)\big)\Big)^{\bot_{E}}.
\end{align*}}That is, $\mathbf{c}\in GRS_{k}(\mathbf{a},\mathbf{v},\infty)^{\bot_{\sigma}}$ if and only if
{\setlength\abovedisplayskip{0.15cm}
\setlength\belowdisplayskip{0.15cm}
\begin{equation*}
\big(v_{1}f(a_{1}),\ldots,v_{n}f(a_{n}),f_{k-1}\big)P_{\tau}^{T}\mathrm{diag}\big(\mu_{1}\pi_{s}(v_{1}),\ldots,\mu_{n}\pi_{s}(v_{n}),\mu_{n+1}\big)
\in\Big(\pi_{s}\big(GRS_{k}(\mathbf{a},\mathbf{1},\infty)\big)\Big)^{\perp_{E}},
\end{equation*}}namely
{\setlength\abovedisplayskip{0.15cm}
\setlength\belowdisplayskip{0.15cm}
\begin{equation*}
\big(w_{\tau^{-1}(1)},\ldots,w_{\tau^{-1}(n)},w_{\tau^{-1}(n+1)}\big)\mathrm{diag}\big(\mu_{1}\pi_{s}(v_{1}),\ldots,\mu_{n}\pi_{s}(v_{n}),\mu_{n+1}\big)
\in\Big(\pi_{s}\big(GRS_{k}(\mathbf{a},\mathbf{1},\infty)\big)\Big)^{\perp_{E}}
\end{equation*}}by using Eq. (\ref{eq2.2}). The desired result is then obtained in terms of Lemma \ref{lemma5.4}. $\hfill\square$
\end{proof}

\begin{remark}\label{remark5.8}
If $P_{\tau}=\mathrm{diag}(P_{\widetilde{\tau}},1)$ for some $n\times n$ permutation matrix $P_{\widetilde{\tau}}$ in Proposition \ref{proposition5.7},
then $\mathbf{c}\in Hull_{\sigma}\big(GRS_{k}(\mathbf{a},\mathbf{v},\infty)\big)$
if and only if there is a polynomial $g(x)\in \mathbb{F}_{q}[x]$ with $\mathrm{deg}\big(g(x)\big)\leq n-k$ such that
\begin{align}\label{eq5.6}
\Big(\mu_{1}\pi_{s}(v_{1})v_{\widetilde{\tau}^{-1}(1)}f\big(a_{\widetilde{\tau}^{-1}(1)}\big),\ldots,
\mu_{n}\pi_{s}(v_{n})v_{\widetilde{\tau}^{-1}(n)}f\big(a_{\widetilde{\tau}^{-1}(n)}\big),\mu_{n+1}f_{k-1}\Big)\notag\\
=\Big(\pi_{s}\big(u_{1}g(a_{1})\big),\ldots,\pi_{s}\big(u_{n}g(a_{n})\big),-\pi_{s}(g_{n-k})\Big).
\end{align}
In particular, if $P_{\widetilde{\tau}}=I_{n}$, then $P_{\tau}=I_{n+1}$ and thus Eq. (\ref{eq5.6}) becomes
\begin{equation}\label{eq5.7}
\big(\mu_{1}\pi_{s}(v_{1})v_{1}f(a_{1}),\ldots,\mu_{n}\pi_{s}(v_{n})v_{n}f(a_{n}),\mu_{n+1}f_{k-1}\big)
=\Big(\pi_{s}\big(u_{1}g(a_{1})\big),\ldots,\pi_{s}\big(u_{n}g(a_{n})\big),-\pi_{s}(g_{n-k})\Big).
\end{equation}
Further, if $D_{\tau}=I_{n+1}$, then Eq. (\ref{eq5.7}) becomes
{\setlength\abovedisplayskip{0.15cm}
\setlength\belowdisplayskip{0.15cm}
\begin{align*}
\big(\pi_{s}(v_{1})v_{1}f(a_{1}),\ldots,\pi_{s}(v_{n})v_{n}f(a_{n}),f_{k-1}\big)
=\Big(\pi_{s}\big(u_{1}g(a_{1})\big),\ldots,\pi_{s}\big(u_{n}g(a_{n})\big),-\pi_{s}(g_{n-k})\Big),
\end{align*}}which contains the following cases:

$\bullet$ For $s=e$, one gets the Euclidean case shown in \cite[Lemma 3]{Chen2018New}:
\begin{align*}
\big(v_{1}^{2}f(a_{1}),\ldots,v_{n}^{2}f(a_{n}),f_{k-1}\big)
=\big(u_{1}g(a_{1}),\ldots,u_{n}g(a_{n}),-g_{n-k}\big).
\end{align*}

$\bullet$ For $s=\frac{e}{2}$ if $e$ is even, one gets the Hermitian case shown in \cite[Lemma 7]{Fang2018Two}:
\begin{align*}
\big(v_{1}^{\sqrt{q}+1}f^{\sqrt{q}}(a_{1}),\ldots,v_{n}^{\sqrt{q}+1}f^{\sqrt{q}}(a_{n}),f_{k-1}^{\sqrt{q}}\big)
=\big(u_{1}g(a_{1}),\ldots,u_{n}g(a_{n}),-g_{n-k}\big).
\end{align*}

$\bullet$ For $s=e-\ell$ with $0\leq \ell\leq e-1$, one gets the $\ell$-Galois case shown in \cite[Proposition II.2]{Cao2021MDS}:
\begin{align}\label{eq5.8}
\big(v_{1}^{p^{\ell}+1}f^{p^{\ell}}(a_{1}),\ldots,v_{n}^{p^{\ell}+1}f^{p^{\ell}}(a_{n}),f_{k-1}^{p^{\ell}}\big)
=\big(u_{1}g(a_{1}),\ldots,u_{n}g(a_{n}),-g_{n-k}\big).
\end{align}
\end{remark}

\subsection{Constructions of $q$-ary MDS codes with new $\ell$-Galois hulls}

As an application of last subsection, we will construct eleven families of $q$-ary MDS codes with new $\ell$-Galois hulls in this subsection.
Note that the parameter $\ell$ of the MDS codes constructed here satisfies $2(e-\ell)\mid e$
while the parameter $\ell$ of those provided in \cite{Cao2021MDS} and \cite{Fang2022New} satisfies $2\ell\mid e$.
Consequently, the eleven families of $q$-ary MDS codes constructed here have new $\ell$-Galois hulls when $\ell\neq \frac{e}{2}$.

Let us outline the key point of our constructions.
After choosing $n$ suitable distinct elements $a_{1},a_{2},\ldots,a_{n}$ as the coordinates of the vector $\mathbf{a}$ in $GRS_{k}(\mathbf{a},\mathbf{v})$
or $GRS_{k}(\mathbf{a},\mathbf{v},\infty)$,
all of the values $u_{1},u_{2},\ldots,u_{n}$ will be determined by Eq. (\ref{eq5.3}).
Then, looking at Eqs. (\ref{eq5.5}) and (\ref{eq5.8}), if we are able to find some $v_{i}$ such that $v_{i}^{p^{\ell}+1}=u_{i}$
(or $v_{i}^{p^{\ell}+1}$ is proportional to $u_{i}$) for $i=1,2,\ldots,n$, then it will be convenient to explore the structure of the polynomial $f(x)$
therein and thus it will be easy to determine the dimensions of $\ell$-Galois hulls of some GRS codes and extended GRS codes.

The following two lemmas will be used in this subsection for the sake of our constructions.

\begin{lemma}\label{lemma5.9}
(\!\!\cite[Lemma III.2]{Cao2021MDS})
Let $q=p^{e}$ with $p$ being an odd prime number and let $0\leq \ell\leq e-1$.
Then for any $u\in \mathbb{F}_{p^{\ell}}^{\ast}$, there exists $v\in \mathbb{F}_{q}^{\ast}$ such that $v^{p^{\ell}+1}=u$ if and only if $2\ell\mid e$.
\end{lemma}

\begin{lemma}\label{lemma5.100}
Let $q=p^{e}$ with $p$ being an odd prime number and let $r$ be an integer with $1\leq r\leq e-1$.
If $(e-r)\mid e$, then $\mathbb{F}_{p^{e-r}}\subseteq \mathbb{F}_{p^{r}}$.
\end{lemma}

\begin{proof}
It suffices to prove that $(e-r)\mid r$. Since $(e-r)\mid e$, there exists some positive integer $x$ such that $e-r=\frac{e}{x}$ and thus $r=e-\frac{e}{x}$.
It is obvious that $\frac{e}{x}\mid (e-\frac{e}{x})$. So $(e-r)\mid r$. $\hfill\square$
\end{proof}

\subsubsection{MDS codes with new Galois hulls related to the norm mapping from $\mathbb{F}_{q}^{\ast}$ to $\mathbb{F}_{p^{e-\ell}}^{\ast}$}

For $q=p^{e}$ with $0\leq \ell\leq e-1$, assume that $(e-\ell)\mid e$.
Consider the \emph{norm mapping} $\mathrm{Nr}:\ \mathbb{F}_{q}^{\ast}\rightarrow\mathbb{F}_{p^{e-\ell}}^{\ast}$
with $\mathrm{Nr}(x):=x^{\frac{q-1}{p^{e-\ell}-1}}$.
Denote $\mathbb{F}_{p^{e-\ell}}^{\ast}=\{b_{1},b_{2},\ldots,b_{p^{e-\ell}-1}\}$. For $1\leq t\leq p^{e-\ell}-1$, denote
{\setlength\abovedisplayskip{0.15cm}
\setlength\belowdisplayskip{0.15cm}
\begin{equation}\label{eq5.9}
\mathcal{N}=\bigcup_{i=1}^{t}N_{i}=\{a_{1},a_{2},\ldots,a_{n}\},
\end{equation}}where $N_{i}$ is defined as
$N_{i}=\{x\in\mathbb{F}_{q}^{\ast}|\ \mathrm{Nr}(x)=b_{i}\}$.
Then, it is verified that $|N_{i}|=\frac{q-1}{p^{e-\ell}-1}$ for each $i$ and
$N_{i}\cap N_{j}=\emptyset$ hold for all $i\neq j$. Thus $n=\frac{t(q-1)}{p^{e-\ell}-1}$ with $1\leq t\leq p^{e-\ell}-1$.
Further, by \cite[Lemma III.1]{Cao2021MDS} one can verify that $a_{i}^{-1}u_{i}\in \mathbb{F}_{p^{e-\ell}}^{\ast}$ for $i=1,2,\ldots,n$,
where $u_{i}$ is defined by Eq. (\ref{eq5.3}).

We now construct the following three families of $q$-ary MDS codes with new $\ell$-Galois hulls.

\begin{theorem}\label{theorem5.11}
Let $q=p^{e}$ with $p$ being an odd prime number. Assume $2(e-\ell)\mid e$.
Let $n=\frac{t(q-1)}{p^{e-\ell}-1}$ with $1\leq t\leq p^{e-\ell}-1$.
Then,

(1) there exists an $[n,k]_{q}$ MDS code with $h$-dimensional $\ell$-Galois hull for any $1\leq k\leq \lfloor\frac{p^{\ell}+n}{p^{\ell}+1}\rfloor$
and $0\leq h\leq k-1$;

(2) there exists an $[n+1,k]_{q}$ MDS code with $h$-dimensional $\ell$-Galois hull for any $1\leq k\leq \lfloor\frac{p^{\ell}+n}{p^{\ell}+1}\rfloor$
and $0\leq h\leq k$;

(3) there exists an $[n+2,k]_{q}$ MDS code with $h$-dimensional $\ell$-Galois hull for any $1\leq k\leq \lfloor\frac{p^{\ell}+n}{p^{\ell}+1}\rfloor$
and $0\leq h\leq k-1$.
\end{theorem}

\begin{proof}
Let $a_{1},a_{2},\ldots,a_{n}$ be defined by Eq. (\ref{eq5.9}).
For each $1\leq i\leq n$, by $a_{i}^{-1}u_{i}\in \mathbb{F}_{p^{e-\ell}}^{\ast}$ and Lemma \ref{lemma5.100},
we have $a_{i}^{-1}u_{i}\in \mathbb{F}_{p^{\ell}}^{\ast}$ and thus $a_{i}^{-1}u_{i}=(a_{i}^{-1}u_{i})^{p^{\ell}}$.
As $2(e-\ell)\mid e$ and by Lemma \ref{lemma5.9}, there exists
$v_{i}\in \mathbb{F}_{q}^{\ast}$ such that $v_{i}^{p^{e-\ell}+1}=a_{i}^{-1}u_{i}$.
Hence $v_{i}^{p^{\ell}+1}=(a_{i}^{-1}u_{i})^{p^{\ell}}=a_{i}^{-1}u_{i}$.
Then, working in an analogous manner as in Theorems III.1, III.2 and III.3 of \cite{Cao2021MDS}, the desired results are obtained. $\hfill\square$
\end{proof}

\begin{remark}\label{remark5.12}
In Table \ref{table1}, we give some examples of $[n,k]_{q}$ MDS codes with $3$-Galois hull from
Theorem \ref{theorem5.11}
for $p=5$, $e=4$, $n=156t+n'$ with $0\leq n'\leq 2$ and $1\leq t\leq 4$,
and $p=7$, $e=4$, $n=400t+n'$ with $0\leq n'\leq 2$ and $1\leq t\leq 6$, respectively.
Since the parameter $\ell$ of the MDS codes with $\ell$-Galois hull in \cite{Cao2021MDS} and \cite{Fang2022New} satisfies $2\ell\mid e$,
our MDS codes with $3$-Galois hull in Table \ref{table1} are not covered by them.
\end{remark}

\begin{table}[!htbp]
\centering	
\small
\setlength{\abovecaptionskip}{0.cm}
\setlength{\belowcaptionskip}{0.2cm}
\caption{Examples of $[n,k]_{q}$ MDS codes with $3$-Galois hull by Theorem \ref{theorem5.11}.}\label{table1}
\vspace{2pt}
\begin{tabular}{cccccccc}
\hline
$\ell$&$p$&$e$&$n$&$k$&$t$&$h$&References\\
\hline
$3$&$5$&$4$&$156t$&$1\leq k\leq \big\lfloor\frac{5^{3}+156t}{5^{3}+1}\big\rfloor$&$1\leq t\leq 4$&$0\leq h\leq k-1$&Theorem \ref{theorem5.11} (1) \\
$3$&$5$&$4$&$156t+1$&$1\leq k\leq \big\lfloor\frac{5^{3}+156t}{5^{3}+1}\big\rfloor$&$1\leq t\leq 4$&$0\leq h\leq k$&Theorem \ref{theorem5.11} (2)\\
$3$&$5$&$4$&$156t+2$&$1\leq k\leq \big\lfloor\frac{5^{3}+156t}{5^{3}+1}\big\rfloor$&$1\leq t\leq 4$&$0\leq h\leq k-1$&Theorem \ref{theorem5.11} (3)\\
\hline
$3$&$7$&$4$&$400t$&$1\leq k\leq \big\lfloor\frac{7^{3}+400t}{7^{3}+1}\big\rfloor$&$1\leq t\leq 6$&$0\leq h\leq k-1$&Theorem \ref{theorem5.11} (1)\\
$3$&$7$&$4$&$400t+1$&$1\leq k\leq \big\lfloor\frac{7^{3}+400t}{7^{3}+1}\big\rfloor$&$1\leq t\leq 6$&$0\leq h\leq k$&Theorem \ref{theorem5.11} (2)\\
$3$&$7$&$4$&$400t+2$&$1\leq k\leq \big\lfloor\frac{7^{3}+400t}{7^{3}+1}\big\rfloor$&$1\leq t\leq 6$&$0\leq h\leq k-1$&Theorem \ref{theorem5.11} (3)\\
\hline
\end{tabular}
\end{table}

\subsubsection{MDS codes with new Galois hulls related to the direct product of two cyclic subgroups}

Denote by $\mathrm{ord}(x)$ the order of the element $x$ in $\mathbb{F}_{q}^{\ast}$.
Let $\xi_{1}:=\alpha^{x_{1}}$ and $\xi_{2}:=\alpha^{x_{2}}$, where $\alpha$ is a primitive element of $\mathbb{F}_{q}$, $x_{1}$ and $x_{2}$ are positive integers.
Let $n=r_{1}r_{2}$ with $1\leq r_{1}\leq \mathrm{ord}(\xi_{1})$ and $r_{2}=\mathrm{ord}(\xi_{2})$.
Denote $\mathcal{R}=\bigcup_{i=1}^{r_{1}} R_{i}=\{a_{1},a_{2},\ldots,a_{n}\}$ with
$R_{i}:=\{\xi_{1}^{i}\xi_{2}^{j}|j=1,\ldots,r_{2}\}$ for $i=1,\ldots,r_{1}$.
If $(q-1)\mid \mathrm{lcm}(x_{1},x_{2})$, then $a_{i}\neq a_{j}$ for all $i\neq j$ by \cite[Lemma III.3]{Cao2021MDS}.
Further, if $\frac{q-1}{p^{e-\ell}-1}\mid x_{1}$ with $(e-\ell)\mid e$,
then $a_{i}^{-1}u_{i}\in \mathbb{F}_{p^{e-\ell}}^{\ast}$ for $i=1,\ldots,n$ by Lemmas III.4 and III.5 of \cite{Cao2021MDS},
where $u_{i}$ is defined by Eq. (\ref{eq5.3}).

Similar to Theorems III.4, III.5 and III.6 of \cite{Cao2021MDS}, we obtain three families of MDS codes with new $\ell$-Galois hulls
in the following theorem. We omit their proofs here.

\begin{theorem}\label{theorem5.13}
Let $q=p^{e}$ with $p$ being an odd prime number. Assume $2(e-\ell)\mid e$, $(q-1)\mid \mathrm{lcm}(x_{1},x_{2})$ and $\frac{q-1}{p^{e-\ell}-1}\mid x_{1}$.
Let $n=\frac{r(q-1)}{\mathrm{gcd}(x_{2},q-1)}$ with $1\leq r\leq \frac{q-1}{\mathrm{gcd}(x_{1},q-1)}$.
Then,

(1) there exists an $[n,k]_{q}$ MDS code with $h$-dimensional $\ell$-Galois hull for any $1\leq k\leq \lfloor\frac{p^{\ell}+n}{p^{\ell}+1}\rfloor$
and $0\leq h\leq k-1$;

(2) there exists an $[n+1,k]_{q}$ MDS code with $h$-dimensional $\ell$-Galois hull for any
$1\leq k\leq \lfloor\frac{p^{\ell}+n}{p^{\ell}+1}\rfloor$ and $0\leq h\leq k$;

(3) there exists an $[n+2,k]_{q}$ MDS code with $h$-dimensional $\ell$-Galois hull for any
$1\leq k\leq \lfloor\frac{p^{\ell}+n}{p^{\ell}+1}\rfloor$ and $0\leq h\leq k-1$.
\end{theorem}

\begin{remark}\label{remark5.14}
In Table \ref{table2}, we give some examples of $[n,k]_{q}$ MDS codes with $5$-Galois hull from
Theorem \ref{theorem5.13}
for $p=5$, $e=6$, $n=279r+n'$ with $0\leq n'\leq 2$ and $1\leq r\leq 4$,
and $p=7$, $e=6$, $n=817r+n'$ with $0\leq n'\leq 2$ and $1\leq r\leq 6$, respectively.
Since the parameter $\ell$ of the MDS codes with $\ell$-Galois hull in \cite{Cao2021MDS} and \cite{Fang2022New} satisfies $2\ell\mid e$,
our MDS codes with $5$-Galois hull in Table \ref{table2} are not covered by them.
\end{remark}

\begin{table}[!htbp]
\centering	
\small
\setlength{\abovecaptionskip}{0.cm}
\setlength{\belowcaptionskip}{0.2cm}
\caption{Examples of $[n,k]_{q}$ MDS codes with $5$-Galois hull by Theorem \ref{theorem5.13}.}\label{table2}
\vspace{2pt}
\begin{tabular}{cccccccccc}
\hline
$\ell$&$p$&$e$&$x_{1}$&$x_{2}$&$n$&$k$&$r$&$h$&References\\
\hline
$5$&$5$&$6$&$3906$&$56$&$279r$&$1\leq k\leq \big\lfloor\frac{5^{5}+279r}{5^{5}+1}\big\rfloor$&$1\leq r\leq 4$&$0\leq h\leq k-1$&Theorem \ref{theorem5.13} (1) \\
$5$&$5$&$6$&$3906$&$56$&$279r+1$&$1\leq k\leq \big\lfloor\frac{5^{5}+279r}{5^{5}+1}\big\rfloor$&$1\leq r\leq 4$&$0\leq h\leq k$&Theorem \ref{theorem5.13} (2)\\
$5$&$5$&$6$&$3906$&$56$&$279r+2$&$1\leq k\leq \big\lfloor\frac{5^{5}+279r}{5^{5}+1}\big\rfloor$&$1\leq r\leq 4$&$0\leq h\leq k-1$&Theorem \ref{theorem5.13} (3)\\
\hline
$5$&$7$&$6$&$19608$&$144$&$817r$&$1\leq k\leq \big\lfloor\frac{7^{5}+817r}{7^{5}+1}\big\rfloor$&$1\leq r\leq 6$&$0\leq h\leq k-1$&Theorem \ref{theorem5.13} (1)\\
$5$&$7$&$6$&$19608$&$144$&$817r+1$&$1\leq k\leq \big\lfloor\frac{7^{5}+817r}{7^{5}+1}\big\rfloor$&$1\leq r\leq 6$&$0\leq h\leq k$&Theorem \ref{theorem5.13} (2)\\
$5$&$7$&$6$&$19608$&$144$&$817r+2$&$1\leq k\leq \big\lfloor\frac{7^{5}+817r}{7^{5}+1}\big\rfloor$&$1\leq r\leq 6$&$0\leq h\leq k-1$&Theorem \ref{theorem5.13} (3)\\
\hline
\end{tabular}
\end{table}

\subsubsection{MDS codes with new Galois hulls related to the coset decomposition of a cyclic group}

For a prime power $q=p^{e}$, assume that $(e-\ell)\mid e$ and set $y:=\frac{q-1}{p^{e-\ell}-1}$. Let $m\mid(q-1)$, then
$m$ can be written as $m=m_{1}m_{2}$ with $m_{1}=\frac{m}{\mathrm{gcd}(m,y)}$ and $m_{2}=\mbox {gcd}(m,y)$.
Let $\mathbb{F}_{q}^{\ast}=\langle \alpha \rangle$. Define $H:=\langle \vartheta_{1} \rangle$ with $\vartheta_{1}:=\alpha^{\frac{q-1}{m}}$ and
$G:=\langle \vartheta_{2} \rangle$ with $\vartheta_{2}:=\alpha^{\frac{y}{m_{2}}}$.
Then $\mbox {ord}(H)=m$ and $\mbox {ord}(G)=(p^{e-\ell}-1)m_{2}$.
Since $m_{2}=\mathrm{gcd}(m,y)$, we have $\mbox {gcd}(m_{1},\frac{y}{m_{2}})=1$. Combining it with $m_{1}\mid \frac{q-1}{m_{2}}$,
we obtain $m_{1}\mid (p^{e-\ell}-1)$, which implies that $H$ is a subgroup of $G$.
Thus the left coset decomposition of $G$ with respect to $H$ can be written as $G=\bigcup_{i=1}^{\frac{p^{e-\ell}-1}{m_{1}}} \eta_{i}H$,
where $\eta_{i}$ is the left coset representative of $G/H$ for $i=1,2,\ldots,\frac{p^{e-\ell}-1}{m_{1}}$.

Let $n=rm$ with $1\leq r\leq \frac{p^{e-\ell}-1}{m_{1}}$. Denote
$\mathcal{H}=\bigcup_{i=1}^{r} \eta_{i}H=\{a_{1},a_{2},\ldots,a_{n}\}$.
It follows from \cite[Lemma III.6]{Cao2021MDS} that $a_{i}^{-1}u_{i}\in \mathbb{F}_{p^{e-\ell}}^{\ast}$ for $i=1,2,\ldots,n$,
where $u_{i}$ is defined by Eq. (\ref{eq5.3}).

\vspace{5pt}
Similar to Theorems III.7, III.8 and III.9 of \cite{Cao2021MDS}, we obtain three families of MDS codes with new $\ell$-Galois hulls
in the following theorem. We omit their proofs here.

\begin{theorem}\label{theorem5.15}
Let $q=p^{e}$ with $p$ being an odd prime number. Assume $2(e-\ell)\mid e$ and $m\mid (q-1)$.
Let $n=rm$ with $1\leq r\leq \frac{p^{e-\ell}-1}{m_{1}}$ in which $m_{1}=\frac{m}{\mathrm{gcd}(m,y)}$ for $y=\frac{q-1}{p^{e-\ell}-1}$.
Then,

(1) there exists an $[n,k]_{q}$ MDS code with $h$-dimensional $\ell$-Galois hull for any $1\leq k\leq \lfloor\frac{p^{\ell}+n}{p^{\ell}+1}\rfloor$
and $0\leq h\leq k-1$;

(2) there exists an $[n+1,k]_{q}$ MDS code with $h$-dimensional $\ell$-Galois hull for any $1\leq k\leq \lfloor\frac{p^{\ell}+n}{p^{\ell}+1}\rfloor$
and $0\leq h\leq k$;

(3) there exists an $[n+2,k]_{q}$ MDS code with $h$-dimensional $\ell$-Galois hull for any $1\leq k\leq \lfloor\frac{p^{\ell}+n}{p^{\ell}+1}\rfloor$
and $0\leq h\leq k-1$.
\end{theorem}

\begin{remark}\label{remark5.16}
In Table \ref{table3}, we give some examples of $[n,k]_{q}$ MDS codes with $5$-Galois hull and $6$-Galois hull from
Theorem \ref{theorem5.15}
for $p=3$, $e=6$, $n=364r+n'$ with $0\leq n'\leq 2$ and $1\leq r\leq 2$,
and $p=3$, $e=8$, $n=820r+n'$ with $0\leq n'\leq 2$ and $1\leq r\leq 8$, respectively.
Since the parameter $\ell$ of the MDS codes with $\ell$-Galois hull in \cite{Cao2021MDS} and \cite{Fang2022New} satisfies $2\ell\mid e$,
our MDS codes with $5$-Galois hull and $6$-Galois hull in Table \ref{table3} are not covered by them.
\end{remark}

\begin{table}[!htbp]
\centering	
\small
\setlength{\abovecaptionskip}{0.cm}
\setlength{\belowcaptionskip}{0.2cm}
\caption{Examples of $[n,k]_{q}$ MDS codes with $5$-Galois hull and $6$-Galois hull by Theorem \ref{theorem5.15}.}\label{table3}
\vspace{2pt}
\begin{tabular}{ccccccccc}
\hline
$\ell$&$p$&$e$&$m$&$n$&$k$&$r$&$h$&References\\
\hline
$5$&$3$&$6$&$364$&$364r$&$1\leq k\leq \big\lfloor\frac{3^{5}+364r}{3^{5}+1}\big\rfloor$&$1\leq r\leq 2$&$0\leq h\leq k-1$&Theorem \ref{theorem5.15} (1)\\
$5$&$3$&$6$&$364$&$364r+1$&$1\leq k\leq \big\lfloor\frac{3^{5}+364r}{3^{5}+1}\big\rfloor$&$1\leq r\leq 2$&$0\leq h\leq k$&Theorem \ref{theorem5.15} (2)\\
$5$&$3$&$6$&$364$&$364r+2$&$1\leq k\leq \big\lfloor\frac{3^{5}+364r}{3^{5}+1}\big\rfloor$&$1\leq r\leq 2$&$0\leq h\leq k-1$&Theorem \ref{theorem5.15} (3)\\
\hline
$6$&$3$&$8$&$820$&$820r$&$1\leq k\leq \big\lfloor\frac{3^{6}+820r}{3^{6}+1}\big\rfloor$&$1\leq r\leq 8$&$0\leq h\leq k-1$&Theorem \ref{theorem5.15} (1)\\
$6$&$3$&$8$&$820$&$820r+1$&$1\leq k\leq \big\lfloor\frac{3^{6}+820r}{3^{6}+1}\big\rfloor$&$1\leq r\leq 8$&$0\leq h\leq k$&Theorem \ref{theorem5.15} (2)\\
$6$&$3$&$8$&$820$&$820r+2$&$1\leq k\leq \big\lfloor\frac{3^{6}+820r}{3^{6}+1}\big\rfloor$&$1\leq r\leq 8$&$0\leq h\leq k-1$&Theorem \ref{theorem5.15} (3)\\
\hline
\end{tabular}
\end{table}

\subsubsection{MDS codes with new Galois hulls related to an additive subgroup of $\mathbb{F}_{q}$ and its cosets}

For an odd prime power $q=p^{e}$, assume $a\mid e$.
Let $K$ denote a $\mathbb{F}_{p^{a}}$-subspace of $\mathbb{F}_{q}$ of dimension $w$
satisfying $\{0\}\subsetneq K\subsetneq \mathbb{F}_{q}$. Then $1\leq w\leq \frac{e}{a}-1$.
Choose $\eta\in \mathbb{F}_{q}\backslash K$ and
set $\mathbb{F}_{p^{a}}=\{\beta_{1},\beta_{2},\ldots,\beta_{p^{a}}\}$.
For $1\leq t\leq p^{a}$, denote
$\bigcup_{i=1}^{t} K_{i}=\{a_{1},a_{2},\ldots,a_{n}\}$,
where $K_{i}:=K+\beta_{i}\eta$.
Then $n=tp^{aw}$. By \cite[Lemma 3.1]{Fang2020Euclidean}, there exists $\varepsilon\in \mathbb{F}_{q}^{\ast}$
such that $\varepsilon u_{i}\in \mathbb{F}_{p^{a}}^{\ast}$ for each $i$, where $u_{i}$ is defined by Eq. (\ref{eq5.3}).
Assume $a\mid (e-\ell)$ and $2(e-\ell)\mid e$, then $\varepsilon u_{i}\in \mathbb{F}_{p^{e-\ell}}^{\ast}$
and thus $\varepsilon u_{i}\in \mathbb{F}_{p^{\ell}}^{\ast}$ by Lemma \ref{lemma5.100}.
So $\varepsilon u_{i}=(\varepsilon u_{i})^{p^{\ell}}$.
Since $2(e-\ell)\mid e$ and by Lemma \ref{lemma5.9}, we know there exists
$v_{i}\in \mathbb{F}_{q}^{\ast}$ such that $v_{i}^{p^{e-\ell}+1}=\varepsilon u_{i}$.
Hence $v_{i}^{p^{\ell}+1}=(\varepsilon u_{i})^{p^{\ell}}=\varepsilon u_{i}$.

\vspace{6pt}

Similar to Theorems III.10 and III.11 of \cite{Cao2021MDS}, we obtain two families of MDS codes with new $\ell$-Galois hulls
in the following theorem. We omit their proofs here.

\begin{theorem}\label{theorem5.17}
Let $q=p^{e}$ with $p$ being an odd prime number. Assume $2(e-\ell)\mid e$ and $a\mid (e-\ell)$.
Let $n=tp^{aw}$ with $1\leq t\leq p^{a}$ and $1\leq w\leq \frac{e}{a}-1$.
Then,

(1) there exists an $[n,k]_{q}$ MDS code with $h$-dimensional $\ell$-Galois hull for any $1\leq k\leq \lfloor\frac{p^{\ell}+n-1}{p^{\ell}+1}\rfloor$
and $0\leq h\leq k$;

(2) there exists an $[n+1,k]_{q}$ MDS code with $h$-dimensional $\ell$-Galois hull for any $1\leq k\leq \lfloor\frac{p^{\ell}+n-1}{p^{\ell}+1}\rfloor$
and $0\leq h\leq k-1$.
\end{theorem}

\begin{remark}\label{remark5.18}
In Table \ref{table4}, we give some examples of $[n,k]_{q}$ MDS codes with $6$-Galois hull and $9$-Galois hull from
Theorem \ref{theorem5.17}
for $p=5$, $e=8$, $n=5^{2w}t+n'$ with $0\leq n'\leq 1$, $1\leq t\leq 25$ and $1\leq w\leq 3$,
and $p=7$, $e=12$, $n=7^{3w}t+n'$ with $0\leq n'\leq 1$, $1\leq t\leq 343$ and $1\leq w\leq 3$, respectively.
Since the parameter $\ell$ of the MDS codes with $\ell$-Galois hull in \cite{Cao2021MDS} and \cite{Fang2022New} satisfies $2\ell\mid e$,
our MDS codes with $6$-Galois hull and $9$-Galois hull in Table \ref{table4} are not covered by them.
\end{remark}

\begin{table}[!htbp]
\centering	
\small
\setlength{\abovecaptionskip}{0.cm}
\setlength{\belowcaptionskip}{0.2cm}
\caption{Examples of $[n,k]_{q}$ MDS codes with $6$-Galois hull and $9$-Galois hull by Theorem \ref{theorem5.17}.}\label{table4}
\vspace{2pt}
\begin{tabular}{cccccccccc}
\hline
$\ell$&$p$&$e$&$a$&$n$&$k$&$t$&$w$&$h$&References\\
\hline
$6$&$5$&$8$&$2$&$5^{2w}t$&$1\leq k\leq \big\lfloor\frac{5^{6}+5^{2w}t-1}{5^{6}+1}\big\rfloor$&$1\leq t\leq 25$&$1\leq w\leq 3$&$0\leq h\leq k$&Theorem \ref{theorem5.17} (1) \\
$6$&$5$&$8$&$2$&$5^{2w}t+1$&$1\leq k\leq \big\lfloor\frac{5^{6}+5^{2w}t-1}{5^{6}+1}\big\rfloor$&$1\leq t\leq 25$&$1\leq w\leq 3$&$0\leq h\leq k-1$&Theorem \ref{theorem5.17} (2) \\
\hline
$9$&$7$&$12$&$3$&$7^{3w}t$&$1\leq k\leq \big\lfloor\frac{7^{9}+7^{3w}t-1}{7^{9}+1}\big\rfloor$&$1\leq t\leq 343$&$1\leq w\leq 3$&$0\leq h\leq k$&Theorem \ref{theorem5.17} (1)\\
$9$&$7$&$12$&$3$&$7^{3w}t+1$&$1\leq k\leq \big\lfloor\frac{7^{9}+7^{3w}t-1}{7^{9}+1}\big\rfloor$&$1\leq t\leq 343$&$1\leq w\leq 3$&$0\leq h\leq k-1$&Theorem \ref{theorem5.17} (2)\\
\hline
\end{tabular}
\end{table}

\section{Conclusion}\label{section6}

The main contribution of this paper are summarized as follows:
\begin{itemize}
\setlength{\itemsep}{1pt}
\setlength{\parsep}{1pt}
\setlength{\parskip}{1pt}
\item We showed that the dimension of the intersection of two linear codes can be determined by generator matrices of such codes and their $\sigma$ duals
(see Theorem \ref{theorem3.4}).
We showed that the dimension of $\sigma$ hull of a linear code can be determined by a generator matrix of it or its $\sigma$ dual
(see Theorem \ref{theorem3.6}).

\item We characterized $\sigma$ duals, $\sigma$ hulls and intersections of matrix-product codes
(see Theorems \ref{theorem4.4}, \ref{theorem4.6} and \ref{theorem4.10}, respectively).

\item We provided a necessary and sufficient condition under which any codeword of a GRS code or an extended GRS code
is contained in its $\sigma$ dual (see Propositions \ref{proposition5.5} and \ref{proposition5.7}).

\item As an application, we constructed eleven families of $q$-ary MDS codes with new $\ell$-Galois hulls satisfying $2(e-\ell)\mid e$
(see Theorems \ref{theorem5.11}, \ref{theorem5.13}, \ref{theorem5.15} and \ref{theorem5.17}).
These codes are not covered by the latest papers \cite{Cao2021MDS} and \cite{Fang2022New} when $\ell\neq\frac{e}{2}$.
\end{itemize}

We remark that an interesting problem for the future is to construct MDS EAQECCs by using the dimension of $\sigma$ hulls of linear codes.
Besides, it will be interesting to find MDS codes with $\ell$-Galois hulls satisfying $2\ell\nmid e$ and $2(e-\ell)\nmid e$.

\section*{Data availability statement}

The datasets generated or analyzed during this study are available from the corresponding author on reasonable request.

\footnotesize{
\bibliographystyle{plain}
\phantomsection
\addcontentsline{toc}{section}{References}
\bibliography{2023.3.18sigma}
}

\end{document}